\documentclass[3p]{elsart}
\usepackage{graphicx}
\usepackage{multirow} 
\usepackage{amssymb}

\usepackage[switch]{lineno}

\newcommand{\ZHAireS}{\mbox{ZHA\scriptsize{${\textrm{IRE}}$}\normalsize{\hspace{.05em}S}}\hspace{.45em}}
\newcommand{\zhaires}{\mbox{ZHA\scriptsize{${\textrm{IRE}}$}\normalsize{\hspace{.05em}S}}}

\begin{document}

\begin{frontmatter}

%The paper title
\title{Coherent Cherenkov radio pulses from hadronic showers up to EeV energies}

%All paper authors
\author[USC]{Jaime Alvarez-Mu\~niz,}
\author[USC]{Washington R. Carvalho Jr.,}
\author[UPL]{Mat\'\i as Tueros,}
\author[USC]{Enrique Zas}

\address[USC]{Depto. de F\'\i sica de Part\'\i culas
\& Instituto Galego de F\'\i sica de Altas Enerx\'\i as,
Universidade de Santiago de Compostela, 15782 Santiago
de Compostela, Spain}

\address[UPL]{Depto. de F\'\i sica, Facultad de Ciencias Exactas,\\ 
Universidad Nacional de La Plata, Argentina}

\begin{abstract}

The Cherenkov radio pulse emitted by hadronic showers of energies in the EeV
range in ice is calculated for the first time using  full three dimensional simulations of both shower development and the coherent radio pulse emitted as the excess charge develops in the shower.
A Monte Carlo, \zhaires, has been developed for this purpose combining the
high energy hadronic interaction capabilities of AIRES, and the dense media propagation capabilities of TIERRAS, 
with the precise low
energy tracking and specific algorithms developed to calculate the radio
emission in ZHS. 
A thinning technique is implemented to allow the 
simulation of radio pulses induced by showers up to 10 EeV in ice. 
The code is validated comparing the results for electromagnetic and hadronic showers to 
those obtained with GEANT4 and ZHS codes. 
The contribution to the pulse of other shower particles in addition to 
electrons and positrons, mainly protons, pions and muons, is found to be below
3 $\%$ for 10 PeV and above proton induced showers.
The characteristics of hadronic showers and the corresponding Cherenkov 
frequency spectra are compared with those from purely electromagnetic showers. 
The dependence of the spectra on shower energy and high-energy hadronic model 
is addressed and parameterizations for the radio emission in hadronic showers 
in ice are given for practical applications. 

\end{abstract}

\end{frontmatter}

%%%%%%%%%%%%%%%%%%%%%%%%%%%%%%%%%%%%%%%%%%
\section{Introduction}

The observation of neutrinos of EeV-scale energies is one of the main
priorities in Astroparticle Physics. 
The detection of neutrinos will open a new window to observe parts 
of the Universe shielded by large depths of matter, 
unaccessible to conventional astronomy.
The measurement of diffuse neutrino 
fluxes will provide further clues to the identification of the sources of 
ultra-high energy cosmic ray production, their composition and their
production mechanisms~\cite{becker08}. In addition such detections could have 
important implications for fundamental particle physics. 

A very promising and cost-effective method to detect high-energy neutrino 
interactions was first proposed by G.~A.~Askar'yan in the 1960's
\cite{Askaryan62}. The idea is to detect the Cherenkov radiation at radio
wavelengths generated by the excess number of electrons in the cascade of
particles resulting from a high-energy particle interaction in a dense medium
transparent to radio waves. The development of this excess charge, due to the interactions with 
matter electrons, is often referred to as the 
Askar'yan effect. The effect has been experimentally confirmed in accelerator
experiments at SLAC in media such as sand \cite{Saltzberg_SLAC_sand,Miocinovic_SLAC_sand},
rock salt \cite{Gorham_SLAC_salt} and ice \cite{Gorham_SLAC_ice}, with results
in good agreement with theoretical calculations \cite{alvz06}.
At radio frequencies (MHz-GHz), the emission is coherent and the radiated
power scales with the square of the primary particle energy.
This makes this method very promising for the detection of neutrinos and cosmic rays of the 
highest energies (EeV) \cite{zheleznykh,ZHS91,frichter96}. 

Several experiments have already exploited this technique searching for 
ultra-high energy neutrinos, but no positive detection has been reported so
far. These include 
experiments using the ice cap at the South Pole, such as the
ANITA balloon experiment \cite{ANITAlite,ANITAlong,ANITA_2009_limits} and
the RICE array of antennas buried under the Antarctic ice
\cite{RICE03,RICElimits}. Other arrays are beginning to be constructed, such as 
the Askar'yan Radio Array ARA \cite{ARA} and ARIANNA \cite{ARIANNA}.
There are also a number of projects using the Moon as target for neutrino
interactions, along with radio telescopes on Earth as radiation detectors, 
such as the pioneering Parkes \cite{Parkes96,Parkes07}, GLUE \cite{GLUElimits}, 
Kalyazin \cite{Kalyazin}, NuMoon \cite{NuMoon}, LUNASKA \cite{LUNASKA} 
and RESUN \cite{RESUN}.

The interpretation of data from these experiments requires a detailed
knowledge of the magnitude, angular distribution and frequency-dependence of
the emitted Cherenkov radiation, which can 
then be related back to the properties of the induced cascade. 
This calls for accurate simulations of the Fourier-spectrum 
of coherent Cherenkov radiation from EeV showers in different dense media.
In the past, full simulations of electromagnetic (EM) showers in ice were done
up to PeV energies, using the well-known and well-tested ZHS code \cite{ZHS91,ZHS92}
and different versions of the GEANT code \cite{razzaque04,almvz03}. Also, 
full simulations of hadronic and neutrino-induced showers in ice were carried
out up to 100 TeV with GEANT \cite{McKay_radio}. 
Hybrid Monte Carlo \cite{alvarez_hybrid} and thinning techniques \cite{Hillas} 
were also developed to simulate electromagnetic \cite{alz97,aljpz09,acorne07}, 
hadronic \cite{alz98,acorne07} and neutrino-induced showers \cite{acorne07,alvz99} 
above PeV energies (mainly in ice and water). Above these energies
the LPM effect \cite{LPM,Stanev_LPM} starts to be important in the media under 
experimental consideration \cite{Stanev_LPM,RalstonLPM,alz97}. 
Semi-analytical calculations of the radio-emission have also been performed \cite{alvz00,buniy02}.
Very recently, simulations of not only the Fourier-components of the spectrum but
also of the radio-pulse in the time-domain emitted in electromagnetic showers, 
have been performed with the ZHS code\cite{alrwz10}.

Modeling hadronic showers up to EeV energies is of utmost importance for neutrino detection, 
since they are induced by all neutrino flavors in neutral 
current (NC) interactions, as well as in the hadronic vertex 
of charged current (CC) interactions of muon and tau neutrinos. 
Although at EeV energies only $\sim 20\%$ of the neutrino energy 
is on average carried by the hadronic shower, 
these showers are known to be less affected by the LPM effect \cite{alz98}, 
and their Cherenkov emission does not suffer from the shrinking 
of the Cherenkov cone, which would otherwise reduce the solid angle 
for observation. On the other hand, mixed showers induced in CC 
electron neutrino interactions are composed of a purely electromagnetic shower, 
produced by an energetic electron carrying on average $\sim 80\%$ of the energy 
of the neutrino, and a hadronic shower, initiated by the debris of 
the interacting nucleon. The advantage for neutrino detection of 
these types of interaction is that all the neutrino energy is channeled 
into the resulting shower. However, the observation of the
electromagnetic shower is expected to be very difficult, since the LPM effect 
shrinks the angular distribution of the electric field, reducing 
dramatically the available solid angle for detection.
For these reasons, experiments aiming at detecting neutrinos 
using the radio technique gain most of their acceptance 
from neutrino-induced hadronic showers \cite{RICElimits,ANITAlong,alvarez_icrc09}.

Clearly, accurate simulations of hadronic 
showers with energies above which the LPM effect becomes effective in different dense media 
are needed. In this paper we present the first steps in that direction. 
We have used the well-known AIRES code \cite{aires} in combination with the 
TIERRAS package \cite{TIERRAS} to simulate proton-induced showers in dense
media, such as ice. We have implemented the algorithms to calculate the 
Fourier components of the electric field produced by charged 
particle tracks in the shower,
developed by Zas-Halzen-Stanev in the well-known and well-tested ZHS code \cite{ZHS91,ZHS92}. 
The result is a flexible and powerful code named \zhaires, which allows the simulation
of electromagnetic, hadronic and neutrino-induced showers in a variety of
media, along with their associated coherent radio emission due to the Askar'yan effect.

It is important to stress that in \cite{alz97,alz98} the calculations of the emission properties 
of showers in ice up to EeV energies were only done in an approximate way. They were 
based on the Fourier-transform of simulated longitudinal 
and lateral shower profiles. With \ZHAireS we can obtain the radio-pulse features in a consistent 
manner within a well-tested simulation, computing the electric field emitted by charged 
particles in the shower. In Ref.\cite{McKay_radio}, GEANT simulations were used to calculate 
the radio-pulse properties in the same way as the \ZHAireS code. However these calculations were 
limited to energies below 100 TeV due to GEANT limitations, while with \ZHAireS we can simulate showers 
at EeV energies and above.  
Also, and for the first time, the contribution to the radio pulse of charged pions, muons and protons
is accounted for and quantified,  
and parameterizations of the frequency spectrum of the radio pulse
due to Askar'yan effect in hadronic showers are presented. Previous parameterizations of hadronic showers exist in the literature~\cite{alz98}, but these
are based on less accurate simulations than those presented in this work.

This paper is structured as follows. In Section \ref{newcode} we describe the new
\ZHAireS code and compare its performance and results for electromagnetic
showers with those of the well-tested codes GEANT 4 \cite{GEANT4} and ZHS
\cite{ZHS92,aljpz09}. We also explore thinning techniques in the \ZHAireS code,
essential for the simulation of EeV showers. 
Section \ref{hadronic} is devoted to the simulation of hadronic showers up to EeV energies,
emphasizing their differences with respect to purely electromagnetic 
showers. Section \ref{conclusions} summarizes and concludes the paper. 
In Appendix A, we also give parameterizations of the intensity of the Fourier-spectrum of 
the Cherenkov electric field emitted in hadronic showers, as a function of 
shower energy and observation angle. 

In this paper we concentrate on hadronic showers in ice. 
We defer to future papers the study of coherent radio emission from
EM and neutrino-induced showers. However we show that our results for proton showers can be applied to approximately model the Cherenkov emission 
from hadronic showers induced in high-energy neutrino interactions. 
The code can also be applied to the simulation of radio emission in extensive
air showers~\cite{zhairesarena2010,zhairesair2011}, despite the different emission mechanism \cite{allan71,ARENA08}. 

%%%%%%%%%%%%%%%%%%%%%%%%%%%%%%%%%%%%%%%%%%
\section{Radio emission in electromagnetic showers using the \ZHAireS code}
\label{newcode}

Coherent Cherenkov emission from high-energy showers in a dense medium 
such as ice is due to the excess of electrons over positrons in
the shower. This excess is mainly caused by the low
energy scattering processes, such as Compton, M\o ller and Bhabha scattering, 
that incorporate electrons of the medium into the shower \cite{ZHS92},
as well as to positron annihilation. 
The characteristics of this emission are dependent on the
lateral and longitudinal profiles of the shower\cite{alvz06}. 
The longitudinal profile affects mainly the emission at angles far from the
Cherenkov angle. As the shape of this profile is governed by high energy 
interactions, radio emission is 
strongly affected by the LPM effect, which reduces the probability of high
energy ($E > 1$ PeV in ice) EM interactions. 
The lateral profile affects the features of the spectrum close to the Cherenkov angle at
frequencies typically above $\sim 1$ GHz, and is mainly determined by low energy particle scattering.

Next we describe the key elements of the 
\ZHAireS code, a flexible and powerful Monte Carlo that allows the simulation
of electromagnetic, hadronic and neutrino-induced showers and their associated 
coherent Cherenkov radio emission due to the Askar'yan effect in a variety of media.

%%%%%%%%%%%%%%%%%%%%%%%%%%%%%%%%%%%%%%%%%%
\subsection{The \ZHAireS code}

\ZHAireS is based on the well-known AIRES\footnote{{\bf AIR}shower {\bf E}xtended {\bf S}imulations} 
code \cite{aires} which we have used in combination with the 
TIERRAS package \cite{TIERRAS} to simulate showers in dense media, such as ice. 
We have implemented in AIRES algorithms to calculate the Fourier components of
the electric field produced by  
charged particle tracks in the shower. These algorithms are the same
used in the well-known and well-tested ZHS code, developed by
Zas, Halzen and Stanev \cite{ZHS91,ZHS92}. 

The standard procedure to simulate Cherenkov coherent radio emission due to the Askar'yan
effect in an energetic shower is to divide all charged particle tracks in the 
shower, down to a relatively low threshold energy, 
in small steps which are approximated by straight lines in which particles
travel at constant speed. A space-time position is associated with the end points of
each step, along with the speed and the charge of the particle. This is all
that is needed to calculate the electric field, and to properly take into account
interference effects between all steps (for a detailed discussion and the
relevant equations see \cite{ZHS92,razzaque01}).  
The AIRES package \cite{aires} is a well established Monte Carlo extensive air shower simulation code, 
from which detailed information on the space-time positions of the tracked 
particles can be extracted. This allows the use of AIRES (v2.8.4a) to simulate 
the coherent radio emission due to any type of particle interaction.
However, AIRES by itself can only simulate showers in air, where the
contribution of the Askar'yan mechanism to radio emission is thought to be
small, and other processes, such as geo-synchrotron radiation, are expected to 
be more important \cite{ARENA08}. 
In order to simulate showers in dense, dielectric media, we have used  
the TIERRAS package \cite{TIERRAS}. Originally designed to continue the simulation of
an extensive air shower underground, TIERRAS also enables the simulation 
of particle cascades in other dense media, such as ice, sea water, rock and soil.
TIERRAS accounts for the atomic and mass number dependence of the interaction 
cross-sections and energy-losses of baryons, mesons and leptons, and includes
a detailed treatment of the LPM effect of special relevance for radio emission 
\cite{RalstonLPM,alz97,alz98}.

In this work we concentrate on showers developing in ice, with refraction
index $n=1.78$ and density $\rho=0.924~{\rm g~cm^{-3}}$. 
Using the \ZHAireS + TIERRAS framework, we track $e^\pm$ with energies above 80 keV, and $\pi^\pm$, $\mu^\pm$, $p$ and $\bar{p}$
  above 500 keV, neglecting radio emission for energies below these thresholds
\footnote{Note that due to the isotropically distributed directions of particles
below these thresholds, their contribution to the electric field is 
expected to be very small \cite{alvarez_in_prep}.}. These energy cuts are the lowest available in AIRES. We also do not account for the emission of other charged particles such as kaons, which as will be shown in Section \ref{hadronic} constitute 
a negligible contribution. 

We used the minimum propagation step available in AIRES, which is typically
much smaller than a radiation length. Directly from the AIRES routines we get
the start and end points for each step, along with the charge of the particle 
and propagation time. From these tracks the radio emission is then calculated 
making extensive use of the formula for   
the frequency ($\omega$) spectrum of the electric field $\vec E$, at a
position $\vec x$ in the Fraunhofer limit of large observation distance $R$,
derived in \cite{ZHS91,ZHS92} and reproduced here for self-consistency:  
\begin{equation}
{\vec E}(\omega,{\vec {\rm x}})=
{e \mu_{\rm r}~i \omega \over 2 \pi \epsilon_0 {\rm c}^2}~
{{\rm e}^{i k R } \over R} ~ 
{\rm e}^{i(\omega - \vec k \cdot \vec v) {\rm t}_1}~ 
{\vec v}_{\perp }~ 
\left[{{\rm e}^{i(\omega - \vec k \cdot \vec v) \delta {\rm t}} -1 
                        \over i (\omega - \vec k \cdot \vec v)} \right]
\end{equation}
where $\mu_{\rm r}$ is the relative magnetic permeability of the medium,
$\epsilon_0$ the free space permittivity, $c$ the speed of light in vacuum, 
$k=\omega n/c$ is the wave vector in a medium with refractive index $n$, 
$\vec v$ is the particle velocity, 
${\rm t}_1$ the start time and $\delta {\rm t}$ is the time interval between 
the end points of the track. 

We recall that this equation has been obtained with the following convention for the Fourier transform of the electric field: 
\begin{equation}
\tilde f(\omega)=2\int_{-\infty}^{\infty}{f}(t)~e^{i\omega t}dt
\label{FT}
\end{equation}
where the factor 2 corresponds to an unusual convention (this factor is usually
either 1 or $(2\pi)^{-{1\over2}}$). This does not pose a problem as long as the 
inverse Fourier transform is done consistently (see \cite{alrwz10} for more details).

The necessary division of the 
particle paths into piecewise linear tracks for radio pulse calculations can be performed 
in various ways, with shorter divisions allowing increasingly accurate parameterizations to higher 
frequencies at the expense of computer time \cite{almvz03}. 
In \ZHAireS the splitting is finer than in the default ZHS \cite{alvz00}, roughly 3-4 times more steps,
which is accurate enough for our purposes up to $\sim$10 GHz and even above, i.e. above the frequency 
at which the coherent behavior of the emitted signal is expected to break down in ice \cite{alvz06}. 
Shorter tracks will lead to a significant increase in computing time but at no improvement in accuracy.

Coherent Cherenkov radio emission in showers in dense media is mainly due to 
electrons and positrons with $\sim$MeV kinetic energies. For instance, $50\%$ of
the projected track length due 
to the excess of negative charge in the shower, a quantity known to be
proportional to the coherent electric field, is produced by electrons with kinetic 
energy between 100 keV and $\sim6$ MeV, with little contribution expected
below 100~keV~\cite{ZHS92}. We have explicitely checked this in our simulations, and the results obtained agree with \cite{ZHS92}. 
A very accurate treatment of this keV to MeV kinetic energy range is thus needed to obtain
a precise determination of the radio emission intensity.
However, in AIRES, electrons from knock-on (KO) electron and positron interactions, 
such as Bhabha and M\o ller scattering,
are not tracked explicitly if the kinetic energy of the incident 
$e^\pm$ is below $K_{\rm low}=1-2$ MeV, and a continuous energy loss $dE/dX$
is used instead. This posed a difficulty, since the 
approximation was not accurate enough to describe the sub-MeV electrons,
and also because in order to fully describe the low energy part of the shower electronic component, 
one should lower $K_{\rm low}$ to explicitly simulate knock-on interactions for electrons down to 
$K_{\rm low}=106$ keV. 
These two issues made the parameterization of $dE/dX$ used in AIRES unsuitable for radio applications. 
To solve these problems, we calculate the continuous ionization loss explicitly for 
low energy electrons and positrons integrating the energy loss for secondaries 
of energy below $K_{\rm low}=106$ keV, as done in the ZHS code, and replace the $dE/dX$ parametrization
implemented in AIRES with the actual Bethe-Heitler energy loss formula at low energies. We have checked that the 
$dE/dX$ obtained in this way is in very good agreement with that implemented in GEANT 4 \cite{GEANT4}, 
with maximum relative differences $\sim 1\%$ when the same $K_{\rm low}$ is
used.

%%%%%%%%%%%%%%%%%%%%%%%%%%%%%%%%%%%%%%%%%%
\subsection{Comparison to GEANT4 and ZHS}

In order to validate the \ZHAireS code we have performed several comparisons of its 
output in ice with that produced by the GEANT 4 \cite{GEANT4} and ZHS \cite{ZHS92}
codes. We have concentrated our efforts in comparing the lateral
(perpendicular to shower axis) development of the excess negative charge in
the shower, as well as the track-length due to this excess negative
charge. For these comparisons, we raised the $e^{\pm}$ kinectic energy cuts to $106$ keV, to match the thresholds used in GEANT4 and ZHS.

\begin{figure}
\begin{center}
\scalebox{0.65}{
\includegraphics{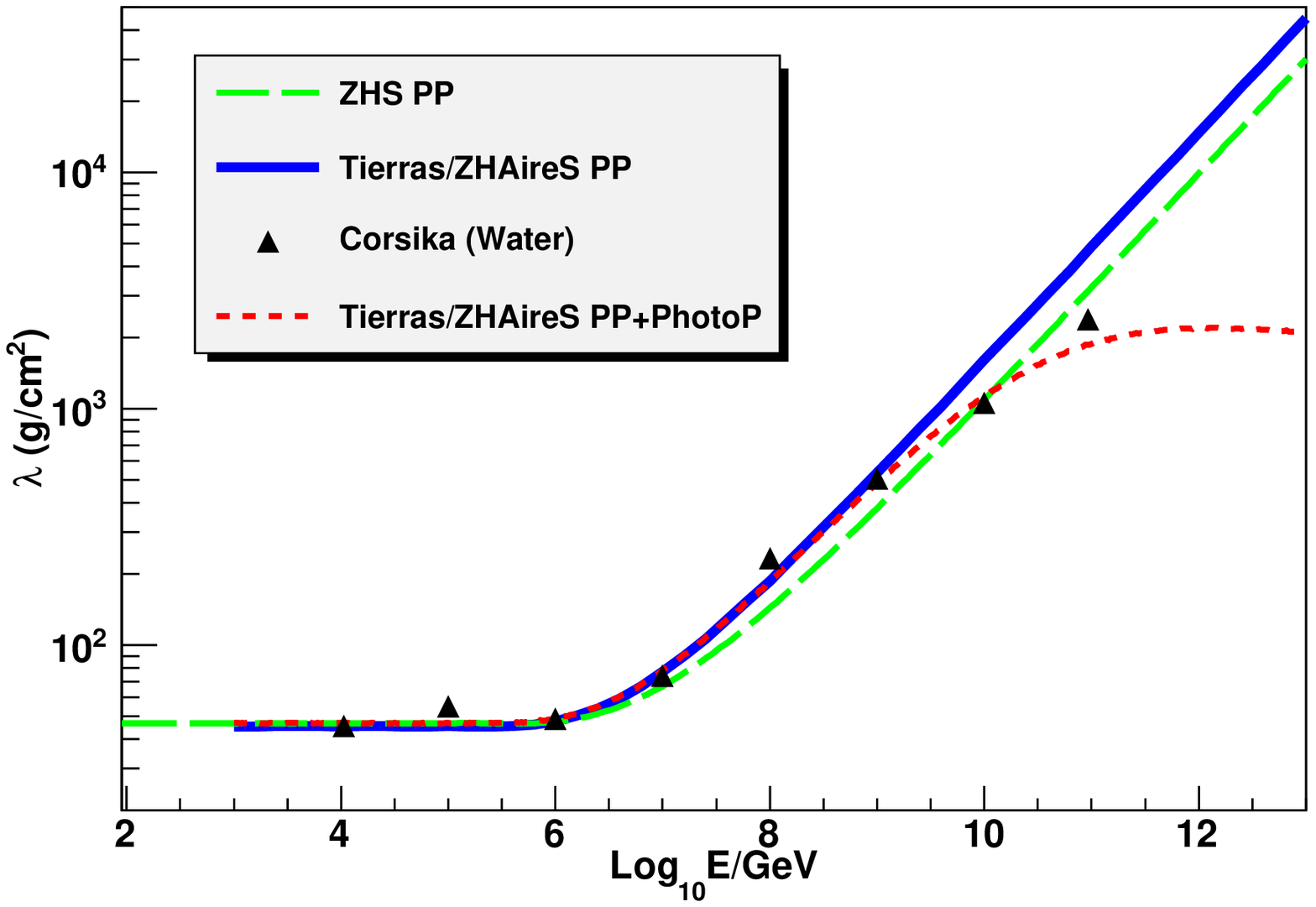}
}
\scalebox{0.65}{
\includegraphics{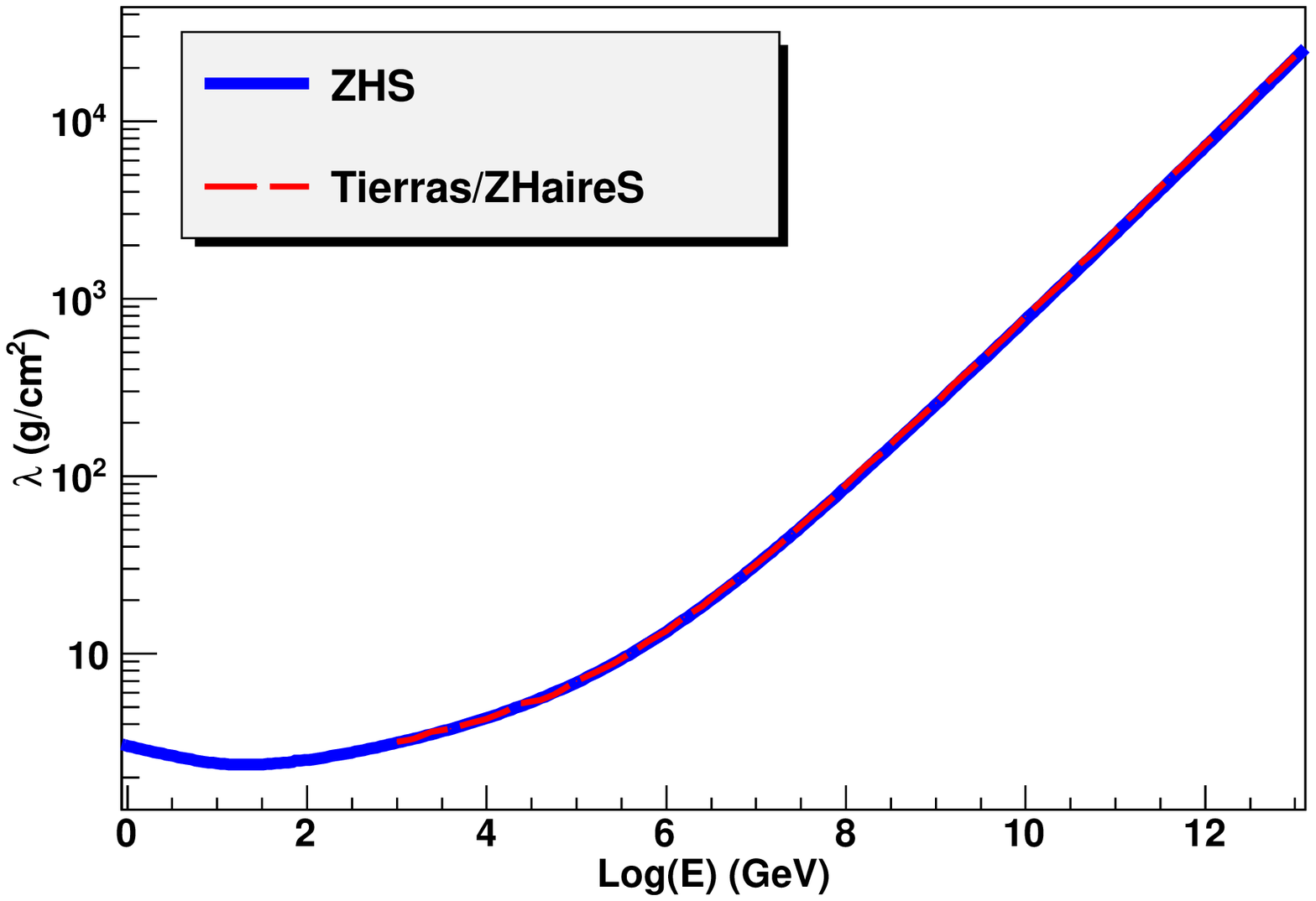}
}
\caption{Top panel: 
Mean free path ($\lambda$) of photons at different energies 
in ZHS (long dashes) and \ZHAireS with (short dashes) and without (solid line) 
photoproduction interactions enabled. Also shown is the interaction
length in CORSIKA \cite{acorne07} with both pair production and photoproduction
enabled (triangles). Note that photoproduction is not accounted for in ZHS.
Bottom panel:
Mean free path of electrons  
of different energies in ZHS (blue solid line) and \ZHAireS (red dashed line
overlapping the blue solid line). 
}
\label{fig:firstinteraction}
\end{center}
\end{figure}

Two relevant quantities affecting significantly the longitudinal development of
an electromagnetic shower are the pair production (PP) and bremsstrahlung 
(BR) cross sections. In order to compare the PP cross-sections in ZHS and
\zhaires \footnote{Note that \ZHAireS uses the same high energy cross-sections
  as the AIRES+TIERRAS framework, including the treatment of the LPM effect.}, 
we obtained the photon mean free path (MFP) at several energies. The result is 
shown in the top panel of Fig.~\ref{fig:firstinteraction}. 
Since photoproduction is not taken into account in ZHS, we disabled 
it temporarily in some \ZHAireS simulations in order to allow a direct 
comparison. Up to 1 PeV, the MFP for PP interactions obtained in \ZHAireS 
is very similar to that in ZHS, with differences $\sim 3\%$, and is equal 
to $(9/7) X_0 \sim 46~{\rm g~cm^{-2}}$, where $X_0=36.08~{\rm g~cm^{-2}}$ 
is the radiation length in ice. Above $\sim~1$ PeV, the energy at which 
the LPM effect starts to produce a decrease in the PP cross section in ice, 
the average interaction depths begin to deviate from each other. For 
photon energies above 100 PeV, the \ZHAireS PP cross-section is around 
$30\%$ lower than the ZHS one, and the LPM effect
turns on at a smaller energy in the \ZHAireS (TIERRAS) code than in 
ZHS. We attribute this difference to details in the implementation of the 
LPM effect in both codes~\cite{cillis,ZHS92,Stanev_LPM}. This will not be investigated
further in this paper, since these differences are mostly irrelevant for the 
hadronic showers and radio pulses discussed below \cite{alz98}. 
Above energies of order 1 EeV, the photoproduction cross-section becomes
larger than the PP cross-section and the interaction length decreases with
respect to the case in which only the LPM PP cross-section 
was accounted for. The energy at which this happens is however 
model-dependent~\cite{new_klein_photoproduction,klein_photoproduction}. The average interaction length
when photoproduction interactions are enabled in \zhaires, 
and in CORSIKA are also shown for comparison. 
 
\begin{figure}
\begin{center}
\scalebox{0.65}{
\includegraphics{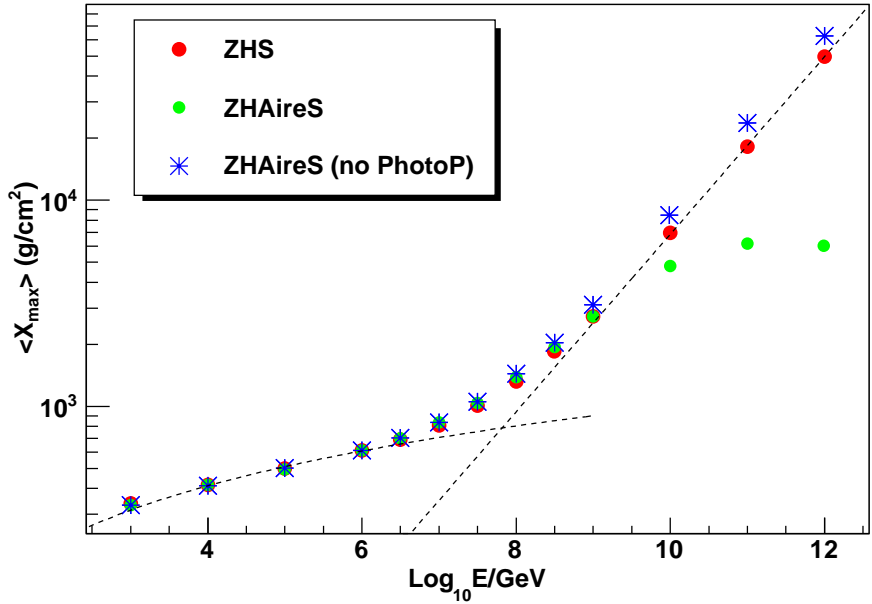}}
\caption{Average depth of shower maximum $X_{\rm max}$ in electron-initiated 
showers simulated with ZHS (dark red dots) and \ZHAireS with 
(light green dots) and without (asterisks) photoproduction enabled. 
The dashed lines, shown just as a reference, are fits to ZHS data: a power law
at high energy and a logarithmic lengthening (elongation rate) at lower energies.}
\label{fig:xmax}
\end{center}
\end{figure}

The difference in the PP cross section is partly responsible for differences 
in some macroscopic shower observables, such as the position $X_{\rm max}$ at
which the number of particles in the shower is maximum. In Fig. \ref{fig:xmax} 
we show  
the average $X_{\rm max}$ in electron-initiated showers simulated with 
ZHS and \ZHAireS (with and without photoproduction interactions enabled). 
The PP cross-section at high energies is lower in \ZHAireS due to the 
difference in the strength of the LPM effect. This translates into an up 
to $\sim 20\%$ larger average $X_{\rm max}$ in \zhaires. If we disable the 
LPM effect and photoreactions in both ZHS and
\zhaires, the elongation rate $\partial X_{\rm max}/\partial \log_{10}E$ 
becomes a constant, as expected, and we obtained a value of 
$81~{\rm g~cm^{-2}}$ in both cases.

In the bottom panel of Fig.~\ref{fig:firstinteraction} we also compare the
mean free path for bremsstrahlung of electrons in \ZHAireS and ZHS simulations which agree to 
better than $\sim 5\%$. 

\begin{figure}
\begin{center}
\scalebox{0.65}{
\includegraphics{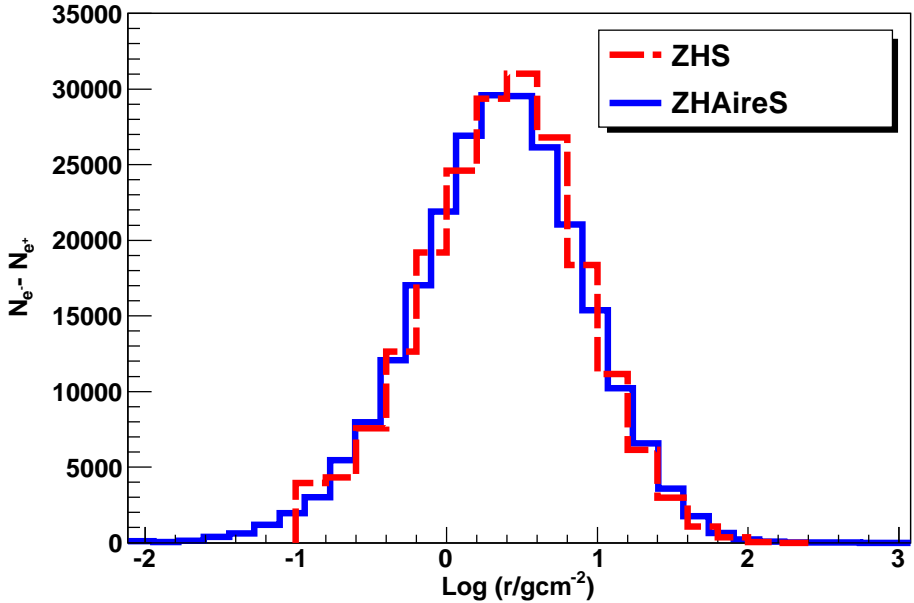}}
\caption{Distribution of the excess negative charge as a function of the 
distance to the shower axis, obtained from simulations of 1 PeV 
electron-induced showers in ice using ZHS (red dashed line)  and \ZHAireS (blue solid line). The lateral distribution refers to a depth around 
$X_{\rm max}\sim~600~{\rm g~cm^{-2}}$. See text for discussion.}
\label{fig:lateral}
\end{center}
\end{figure}

In Fig.~\ref{fig:lateral} we show the lateral distribution of the excess
negative charge, at a depth around shower maximum, obtained from simulations
of 1 PeV electron-induced showers in ice, performed using \ZHAireS and ZHS.
The lateral distribution is a relevant quantity that mainly affects the 
frequency spectrum of the field at high frequencies, typically above $\sim$ 1 
GHz. The agreement between both codes is rather good\footnote{The differences 
near shower axis are mostly due to the fact that ZHS by default includes all 
particles with $r<0.1~{\rm g~cm^{-2}}$ in the first bin.} 

In Table \ref{tab:tracks} we show the average track lengths obtained from 100
simulations of 100 GeV electron-induced showers with \ZHAireS and ZHS, and
compare them with GEANT results taken from \cite{almvz03}. 
The total track length is defined as the sum of all electron and positron
tracks in the shower. The total projected track length 
is the sum of all the projections of these tracks onto the shower axis. The
excess projected track length is the difference
between the sum of all electron track projections minus the sum of all
positron track projections. This is the most relevant quantity that determines
the normalization of the electric field spectrum,
since electrons and positrons contribute to the field with opposite signs.  
The relative differences between the total track length and excess projected 
track length as obtained in \ZHAireS and ZHS are $4.1\%$ and $0.9\%$, 
respectively. The corresponding differences between \ZHAireS and GEANT4 are 
$3.6\%$ and $1.0\%$, respectively\footnote{Relative differences are 
calculated using [\ZHAireS - ZHS]/ZHS and [\ZHAireS - GEANT4]/GEANT4.}.  

\begin{table*}[!htb]
\begin{tabular}{ l c c c c } \hline
{MC Code}&  {\ZHAireS}&  {GEANT 3.21}&  {GEANT 4}& {ZHS}\\ \hline\hline
{Total Track [m]}& 566.8& 577.9&  587.9&  591.1\\  
{Total Projected [m]}& 459.3&  450.0&  453.2& 480.9\\ 
{Excess Projected [m]}& 123.9&  123.5& 122.7& 125.1\\ 
{Excess Projected/Total}& 0.219&  0.214& 0.209& 0.212 \\ \hline
\end{tabular}
\caption{Comparison between average track lengths obtained from simulations 
of electron-induced showers of $E_0=$100 GeV in ice, using \zhaires, GEANT 3.21, 
GEANT 4 and the latest version of ZHS \cite{aljpz09}. The results of \ZHAireS and ZHS  
are the average of 100 simulated showers each, while the results of the GEANT 3.21 and GEANT 4 codes
are taken from \cite{almvz03}. The total excess track length is 193.4 m in \ZHAireS and 
200.3 m in ZHS. Corresponding total excess track lengths for GEANT are not available. 
Note that the results shown here for 
the ZHS code give better agreement than reported in~\cite{almvz03}, due to  
a minor correction in the bremsstrahlung terms proportional to the density of matter 
electrons. 
}
\label{tab:tracks}
\end{table*}

\begin{figure}
\begin{center}
\scalebox{0.65}{
\includegraphics{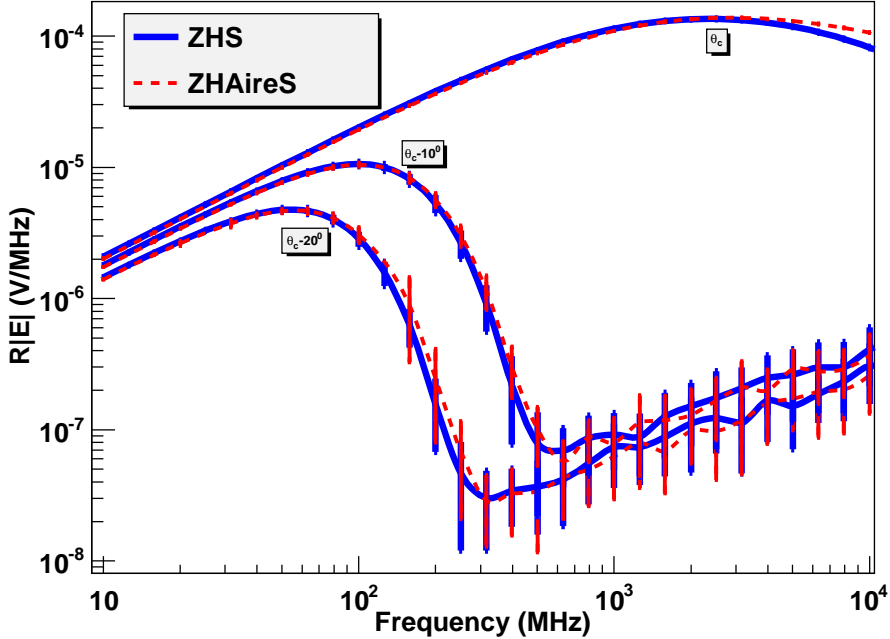}}
\caption{
Average frequency spectrum of the Cherenkov radiation obtained
in simulations of 20 electron-induced showers with primary energy $E_0=1$
PeV, using ZHS (blue solid line) and \ZHAireS (red dashed line). The spectrum 
is shown at three observation angles with respect to the shower axis. 
The RMS of the 20 simulated showers is also shown.}
\label{fig:e1pev-zhs-zhaires}
\end{center}
\end{figure}

In Fig. \ref{fig:e1pev-zhs-zhaires} we compare the average frequency spectra 
of the Cherenkov radiation obtained from 20 ZHS simulations of 
electron-induced showers of primary energy $E_0=1$ PeV, with the 
corresponding average spectrum obtained with the \ZHAireS code. 
One can see a very good agreement below the cut-off frequency in the
spectrum (the frequency at which the electric field peaks before decreasing). 
Beyond the cut-off frequency the agreement worsens. At the Cherenkov 
angle and frequencies above about 3 GHz, the electric field is sensitive to 
the fine details of the shower, in particular to the different splitting of 
the charged particle tracks in ZHS and \ZHAireS codes. 
The finer splitting in the \ZHAireS code (3-4 times more steps)
compared to ZHS \cite{aljpz09} in which the electric field
is calculated within the approximation ``b" \cite{alvz00},  leads to 
a higher spectrum in the GHz frequency range, when compared to ZHS. 
The effect of the splitting on the high-frequency Fourier components of the 
electric field was demonstrated numerically in \cite{alvz00}. At higher energies, typically above 1 EeV or so photoproduction interactions start to play an important role in
ice (see Fig. \ref{fig:xmax} and Ref. \cite{Klein-Gerhardt}). While in the ZHS
code only EM processes are accounted for we expect the results of the ZHAireS
and ZHS code for radio emission in EM showers to differ. This topic will be studied in a forthcoming paper \cite{alvarez_in_prep}.

%%%%%%%%%%%%%%%%%%%%%%%%%%%%%%%%%%%%%%%%%%
\subsection{Thinning in \ZHAireS}
\label{thinning}

In order to make possible the simulation of showers above PeV energies, 
AIRES uses thinning algorithms \cite{Hillas,aires} 
to reduce computing time significantly. The basic idea of thinning 
is to follow only a small, representative fraction of the particles 
in a shower, and assign to each tracked particle 
a corresponding weight to compensate for the rejected ones \cite{Hillas}. 

Based on the properties of radio emission, a study 
was made in \cite{aljpz09} of a thinning technique for calculations of 
radio emission in EM showers.
The conclusion was that applying a thinning algorithm similar to the original
Hillas algorithm \cite{Hillas}, but with two thinning thresholds instead of one, i.e., 
thinning only particles with energies between $E_{\rm min}$ 
and $E_{\rm max}$, produces an accurate representation of the coherent 
Cherenkov radio-emission in electromagnetic showers, while at the same 
time keeps computing time to an acceptable value. The thinning parameters
obtained in \cite{aljpz09} were $E_{\rm min}\sim~100$ MeV $-$ 1 GeV and 
$E_{\rm max}\sim~(10^{-4}-10^{-5})~E_0$, where $E_0$ is shower energy. These parameters
were obtained as a compromise between accuracy (smaller than $\sim 10\%$) and CPU time. 
This algorithm is now implemented in the ZHS code.

The \ZHAireS code uses the same thinning algorithm as AIRES, with an energy
threshold $E_{\rm th}$, equivalent to $E_{\rm max}$ in ZHS, and a so-called 
weight factor $W_f$ \cite{aires}, which sets a maximum weight allowed 
in the simulation. 
It is important to remark that the weight factor is not the maximum weight of the particles
in the simulation, and that the maximum weight increases linearly with shower energy \cite{aires}.
By fine-tuning $E_{\rm th}$ and
$W_f$, one can make $W_f$ work in a similar fashion as $E_{\rm min}$, thus
making the \ZHAireS thinning algorithm work in a similar way as the one 
in ZHS. 

\begin{figure}
\begin{center}
\scalebox{0.65}{
\includegraphics{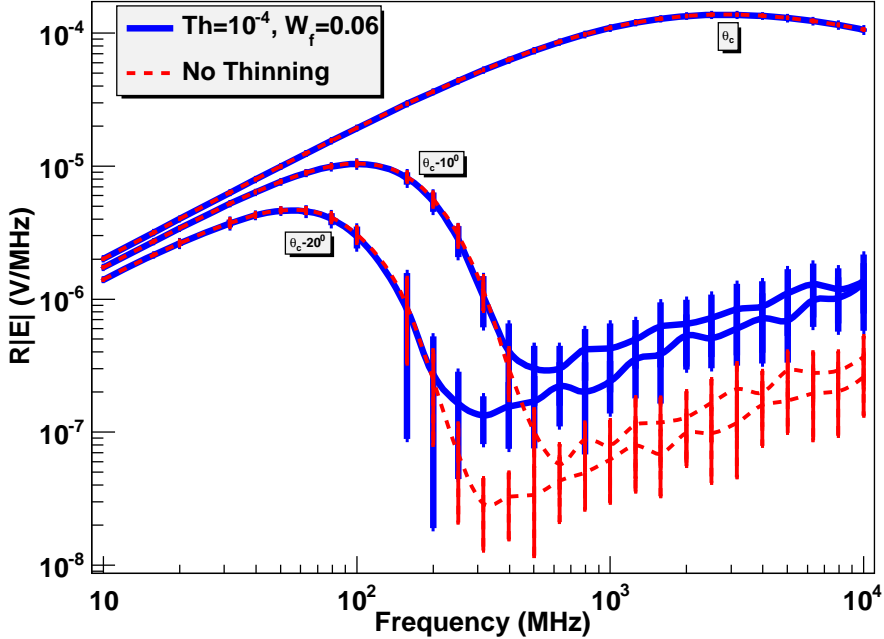}}
\caption{Average frequency spectrum of the Cherenkov radiation generated by
  electron-induced showers with primary energy $E_0=1$ PeV, obtained from
  20 un-thinned simulations (red dashed line) and 20 thinned simulations 
(blue solid line), using the \ZHAireS code. The spectrum is shown at three 
observation angles with respect to shower axis. The RMS of the 20 simulated 
showers is also shown. The parameters used in the thinned
  simulation were $E_{\rm th}=10^{-4}~E_0$ and $W_{f}=0.06$.}
\label{fig:e1pev-nothin-th-4-0.06wf}
\end{center}
\end{figure}

The key point is to select thinning parameters 
($E_{\rm th}$ and $W_f$, in the case of the 
\ZHAireS code), so that the computing time is minimized, while 
maximizing the agreement between full
simulations, i.e., those performed following all particles, and the thinned 
ones. For that purpose we have compared full simulations
of 20 showers at 1 PeV with thinned simulations using different choices 
of $E_{\rm th}$ and $W_f$, in a similar fashion to 
what was done in \cite{aljpz09}. A good compromise between accuracy and 
computing time was obtained for the parameters $E_{\rm th}=10^{-4}~E_0$ 
(as in the case of ZHS \cite{aljpz09}) and $W_f=0.06$. An example is shown in 
Fig.~\ref{fig:e1pev-nothin-th-4-0.06wf}, where we plot a comparison 
between 20 thinned simulations of electron-induced showers of energy $1$ PeV
performed with the thinning parameters above, and 20 
full simulations of the same energy, both using the \ZHAireS code. 
One can see that the agreement is very good up to frequencies beyond the 
cutoff frequency in the spectra, which depends on the observation angle. In
this region, the difference between the average electric field in thinned 
and full simulations is always below $4\%$, and the difference 
in the RMS of the simulated showers normalized to the average value, i.e., 
$(\sigma_{E}^{\rm thin}-\sigma_{E }^{\rm unthin})/ \langle |\vec{E}| \rangle$ 
with $\langle | \vec{E}| \rangle$ the electric field, is always smaller than $\sim 5\%$. However, 
for observation angles away from the Cherenkov angle 
and frequencies well beyond the cutoff in the spectrum, the two 
calculations deviate significantly. 
This is not surprising since the Fourier components are sensitive 
to the fine structure of the shower (even at the individual particle level) 
in this angular and frequency region,
and clearly the thinning algorithm in the \ZHAireS code is unable to 
account for it.  
This is not a problem because in that region 
the electric field strength is typically
a factor 10 to 100 smaller than in the coherent region,
and it is not expected to contribute significantly to the
total emitted power.

\begin{figure}
\begin{center}
\scalebox{0.65}{
\includegraphics{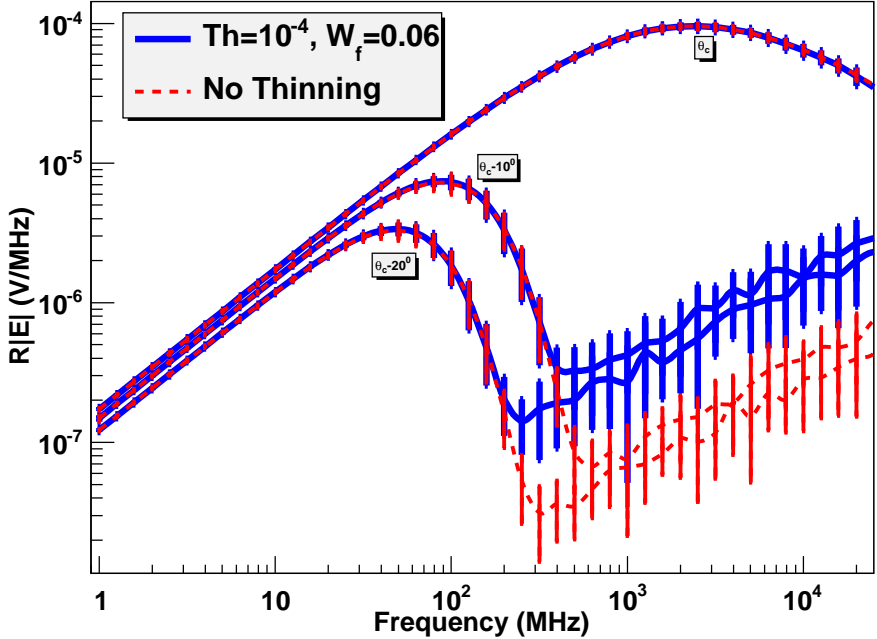}}
\caption{Same as Fig.~\ref{fig:e1pev-nothin-th-4-0.06wf} 
for proton-induced showers.}
\label{fig:p1pev-nothin-th-4-0.06wf}
\end{center}
\end{figure}

At energies above $\sim 1-10$ PeV, simulating showers without thinning 
becomes impractical, since computing time scales linearly with shower 
energy. This makes a direct comparison between
thinned and un-thinned simulations above 1-10 PeV almost impossible. 
Instead, we have compared \ZHAireS thinned simulations, using the same 
parameters described before,  with thinned simulations performed with ZHS, 
using its own thinning parameters ($E_{\rm min}$ and $E_{\rm max}$), 
for which
a comprehensive study of their validity was made \cite{aljpz09}. The agreement
between the two simulations is excellent at all energies, and we adopt the 
thinning parameters ($E_{\rm th}=10^{-4}~E_0,~W_f=0.06$) in our 
subsequent simulations.  

%%%%%%%%%%%%%%%%%%%%%%%%%%%%%%%%%%%%%%%%%%
\section{Radio emission in hadronic showers}
\label{hadronic}

%%%%%%%%%%%%%%%%%%%%%%%%%%%%%%%%%%%%%%%%%%

In this section we study the coherent Cherenkov
radio emission of hadronic showers in ice. Using
the \ZHAireS code, we simulated large samples of 
proton-induced showers up to an energy of $E_0=$10 EeV. 

A first important check to perform is whether the values for the 
thinning parameters $E_{\rm th}$ and $~W_f$, obtained using EM showers
in section \ref{thinning}, are also valid to accurately calculate radio pulses
from hadronic showers. 
For this purpose, in Fig.~\ref{fig:p1pev-nothin-th-4-0.06wf} we compare
20 full simulations of proton-induced showers with 20 showers simulated
using the same thinning parameters obtained for electromagnetic 
showers. The agreement between both simulations is equivalently good except again 
for the frequency region above the region of full coherence.  
The CPU time required remains within reasonable values,  
reinforcing the conclusion that the same thinning algorithms and parameters
can be used safely and provide a good description of both, electromagnetic and 
hadronic showers. 

\subsection{Comparison to GEANT4 simulations}

\begin{figure}
\begin{center}
\scalebox{0.65}{
\includegraphics{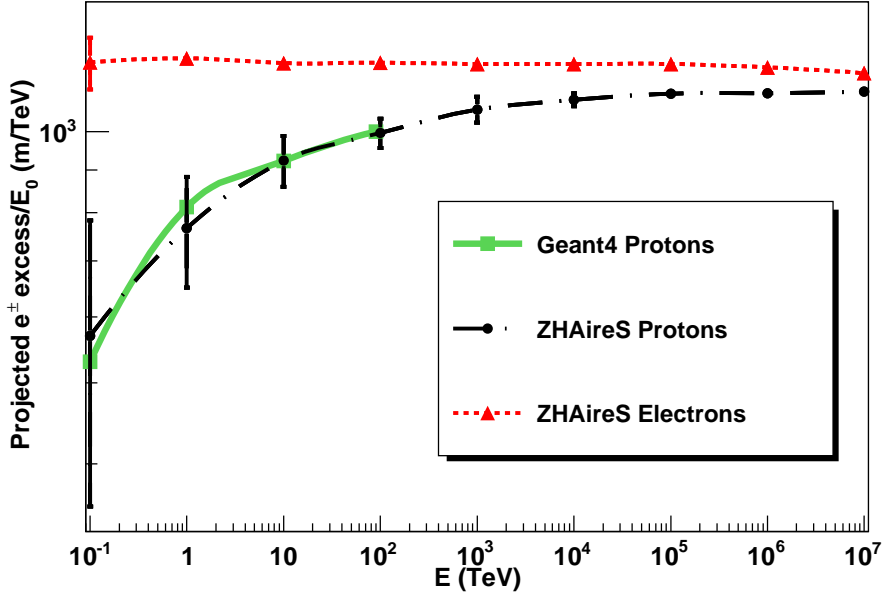}}
\caption{Projected excess electron track length as a function of energy, obtained in 
GEANT4 \cite{McKay_radio} (solid line) and \ZHAireS (dot-dashed line) proton-induced shower 
simulations. For clarity, the track length has been divided by
shower energy. Also shown are the track length for electron-induced showers,
simulated using \zhaires. The GEANT4 results have an uncertainty (not plotted)
of $\sim10\%$ \cite{McKay_radio}.}
\label{fig:mckay}
\end{center}
\end{figure}

In this section we compare our results to 
simulations of proton-induced showers using 
GEANT4, as reported in \cite{McKay_radio}. In 
Fig.~\ref{fig:mckay}, we compare the projected excess track length obtained in 
 simulations of proton-induced showers using GEANT4 and \zhaires. For clarity, 
the projected excess track length has been divided by shower energy $E_0$.
For the purposes of the discussion below, we also show in Fig.~\ref{fig:mckay}
the excess projected track length in the case of electron-induced showers, 
obtained in \zhaires.

GEANT4 simulations of proton showers have only been performed 
up to 90 TeV, while we performed \ZHAireS simulations up to 10 EeV, an energy 5
orders of magnitude higher. One can see that there is a good agreement between
the results of both simulations. The small differences between the track
lengths from both simulations, up to a maximum of $\sim 7\%$, are well within 1
RMS of the \ZHAireS results, which were obtained from 20 simulations for each
primary energy. In the case of the GEANT4 simulations reported in
\cite{McKay_radio}, a RMS $< 10\%$ is quoted, except for 90 TeV, for which
only a single shower was simulated and thus the RMS could not be obtained.

A feature that can be seen in Fig.~\ref{fig:mckay} is that the track length
due to the excess of electrons in hadronic showers approaches that in
electron-induced showers at high energies. This is due to the fact that  the fraction of primary energy going into the electromagnetic component of the shower increases  with shower energy. Assuming energy equipartition between pions, in each interaction a fraction of 1/3 of the incident pion energy would go into the electromagnetic component. Because the medium is dense, the produced charged pions are expected to interact before decaying and hence produce more $\pi^0$s that contribute further to the EM content of the shower. As shower energy increases more particle generations and interactions occur, and hence more energy is transferred to the EM component. Hadronic showers at the highest energies will have a markedly electromagnetic character because of the high energy involved and the medium density which prevents decays of most of the charged pions.

%A feature that can be seen in Fig.~\ref{fig:mckay} is that the track length
%due to the excess of electrons in hadronic showers approaches that in
%electron-induced showers at high energies. This is due to the decrease with
%shower energy of the fraction of the energy going into the hadronic component
%of the shower. As the shower energy increases, more energetic charged pions
%are produced which are more likely to interact than to decay, producing lower
%energy neutral pions, which in turn decay into two photons, increasing the
%energy channeled into the EM component of the shower. This increase of the EM
%energy fraction is also present in atmospheric showers, where it is most
%studied since it is a source of uncertainty for the energy estimates
%\cite{watson-nagano}. Another important difference with respect to
%atmospheric showers is that in ice, the energy at which a charged pion is
%more likely to interact than to decay is roughly 3 orders of magnitude
%smaller, making the EM  energy fraction higher in ice than in air for showers
%of the same energy. 

To further check this interpretation we have explicitly obtained in
\ZHAireS simulations the energy going into the EM component of the shower. The
result of this study shows that the increase with shower energy of the EM
energy fraction is responsible for the increase with energy of the excess electron track in proton-induced
showers shown in
Fig.~\ref{fig:mckay}. Between 1 TeV and 90 TeV, the ratio between the
GEANT4 $e^{\pm}$ track length calculation, as shown in figure \ref{fig:mckay}, and
the EM energy fraction obtained from \ZHAireS simulations varies only slightly,
between 11.6 and 11.8 m/TeV, showing that the projected $e^{\pm}$ excess
track length is proportional to the EM energy fraction of the shower. We have
also checked that the total track length behaves similarly, i.e. it is
proportional to the EM energy fraction, confirming that the behavior seen in figure \ref{fig:mckay} is not an effect
of projecting the track along the shower axis. The EM energy fractions obtained here are also in agreement with those obtained in \cite{alz98}, where a simplified simulation of shower development was performed.

Interestingly, in the case of electron-induced showers, one can see in
Fig.~\ref{fig:mckay} a small deviation from the linear dependence of the
track length with energy. We attribute this behavior to photonuclear
interactions becoming relevant at high energy and contributing to the hadronic
component (i.e. reducing the EM energy fraction), even in electron-induced
showers. This is confirmed by our simulations, as can be seen by the values of
the EM energy fraction in electron induced showers, shown in table \ref{tab:tracks_p}.

\subsection{Radio emission}

The dominant radiation mechanism in hadronic showers is 
the same as in electron-induced ones, namely, emission of coherent
Cherenkov radiation from the excess of electrons
over positrons, i.e., the Askar'yan effect. However, there
are differences between hadronic and electromagnetic showers in the
normalization of the magnitude of the electric field, as well as in some
important features of the frequency spectrum, which stem mainly from the
different spatial distribution  and from the fraction of EM energy in both types of showers \cite{alz98,McKay_radio}. 

\begin{figure}
\begin{center}
\scalebox{0.45}{
\includegraphics{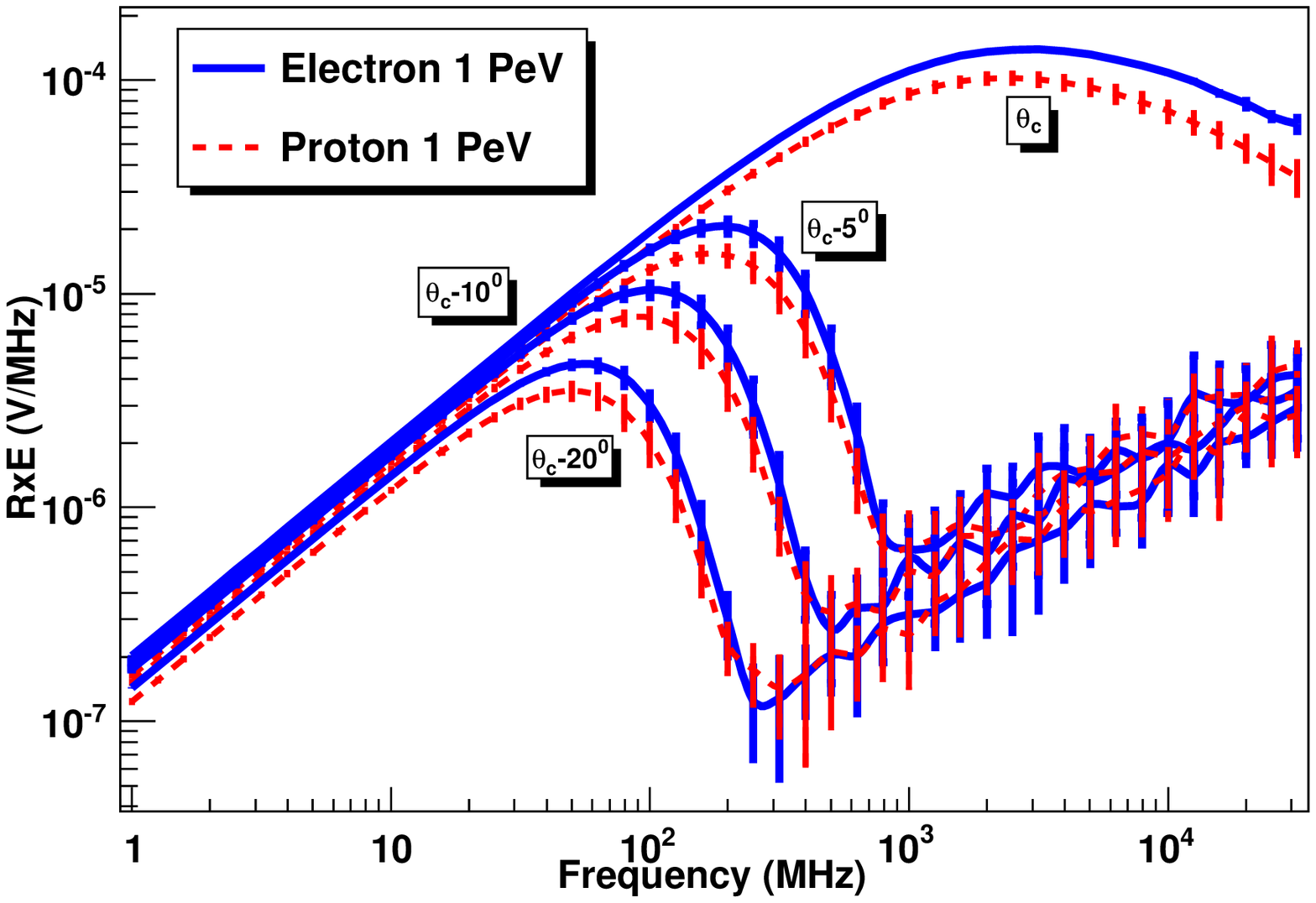}}
\scalebox{0.45}{
\includegraphics{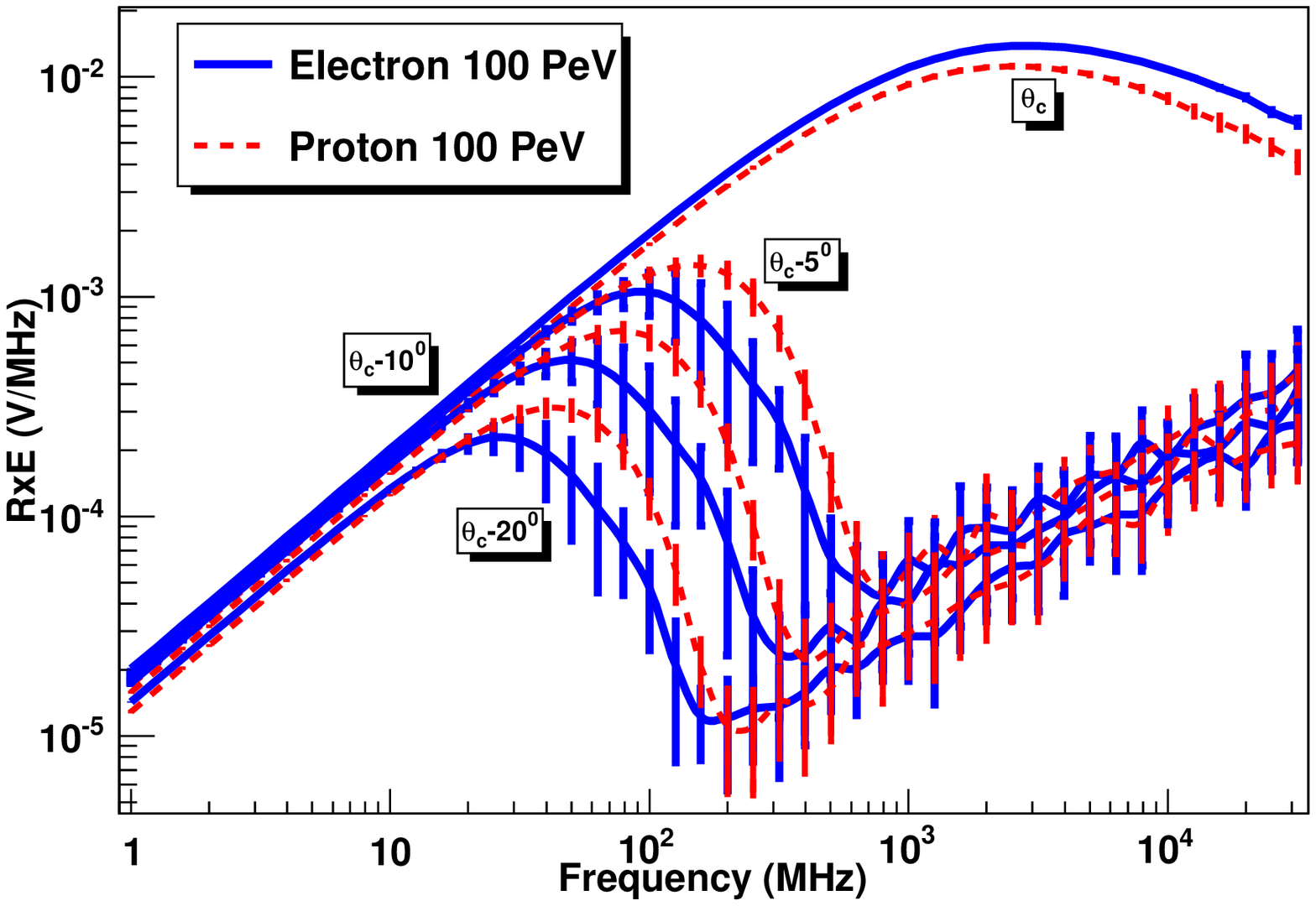}}
\scalebox{0.45}{
\includegraphics{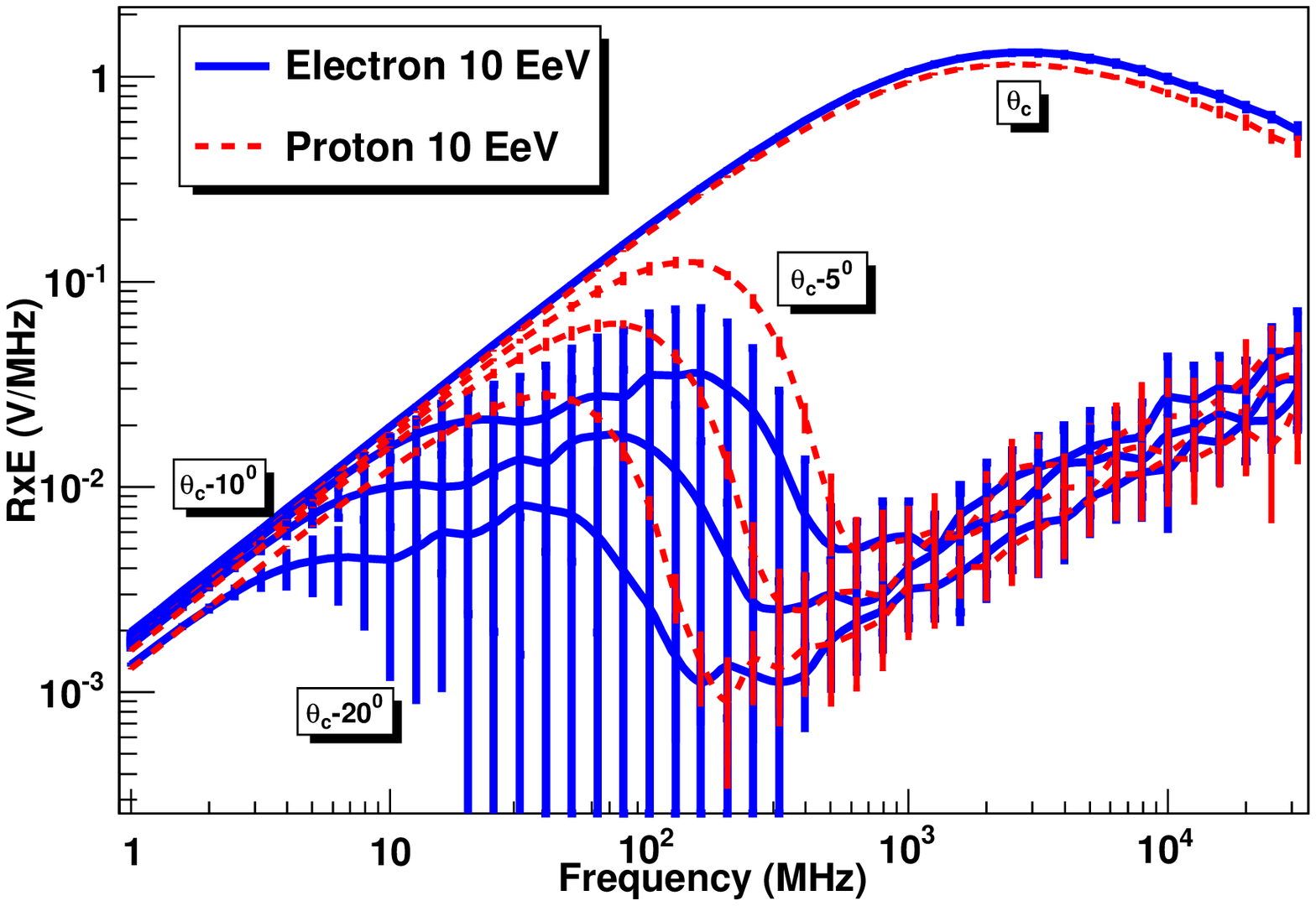}}
\caption{
Average frequency spectrum of the Cherenkov radiation obtained
in \ZHAireS simulations of electron (solid blue line) and proton (dashed red line) showers with primary 
energy $E_0=1$ PeV (top panel), $E_0=100$ PeV (middle panel) and
$E_0=10$ EeV (bottom panel). The spectrum is shown at four observation 
angles with respect to the shower axis. The RMS of the 20 simulated showers is also shown.
}
\label{fig:comp-e-p}
\end{center}
\end{figure}

Fig.\ref{fig:mckay} displays the projected track length excess divided  
by primary shower energy as a function of shower energy for both
electromagnetic and hadronic showers. This ratio is almost constant
  for electromagnetic showers while for hadronic showers it increases with
  energy because the EM energy ratio of hadronic showers also increases with
  energy, as explained before, approaching the value for electromagnetic showers but remaining always smaller. 
This reflects itself in the normalization of the electric
field which is smaller in hadronic showers. This can be appreciated in the 
top panel of Fig~\ref{fig:comp-e-p}, where the Cherenkov spectrum of 
proton and electromagnetic 1 PeV showers are compared. At this energy, the
ratio between the electric fields at 1 MHz and at the Cherenkov angle of the
proton and electron induced showers is 0.864, whilst the ratio between the EM
energy fraction of these showers is very similar, 0.902,
i.e. $\langle |\vec{E}|_{p} \rangle / \langle |\vec{E}|_{e} \rangle \simeq EM^{fraction}_p/EM^{fraction}_e$.  
It can be also noted that as the shower energy increases 
(from top to bottom panels), the electric field produced 
by proton-induced showers approaches that of electromagnetic showers 
at small frequencies, where the coherence is full 
(10 - 100 MHz depending on the observation angle), 
as expected from the results of the excess projected track length 
displayed in Fig~\ref{fig:mckay}. 

A relevant feature of the frequency spectrum is the cut-off
frequency at the Cherenkov angle, which is known
to be inversely proportional to the lateral dimensions of the shower 
\cite{alvz06}. Hadronic showers typically spread over a larger
lateral distance than electromagnetic ones \cite{McKay_radio}, and as a
consequence, the cut-off frequency in the hadronic case is expected to be 
smaller. The Cherenkov spectrum of electron-induced 
showers peaks at $\sim 3$ GHz for observation at the Cherenkov angle, while 
the peak shifts to lower frequencies, around $2.5$ GHz, in proton-induced 
showers. The effect is however difficult to appreciate in 
Fig~\ref{fig:comp-e-p}.     

Another important effect is the shift in the cut-off frequency 
at angles away from the Cherenkov angle, which is known to 
be determined by the longitudinal development of the shower
\cite{alvz06,ZHS92}. Electron-induced showers develop more 
slowly than hadronic ones and hence penetrate more in the 
medium. As a result, for a fixed energy, hadronic showers 
are on average shorter 
in the longitudinal direction (along the shower axis) and 
we expect the cut-off frequency of their spectra, away from 
the Cherenkov angle, to be typically larger than in electron-induced showers. 
This effect is however not visible at energies below the LPM scale 
($\sim$ PeV) (top panel of Fig.~\ref{fig:comp-e-p}), but
it is apparent at energies above which the LPM effect becomes important 
for shower development. 

Hadronic showers are known to be less
affected by the LPM effect than electromagnetic ones \cite{alz98},
the reason is that the production of high energy photons 
through neutral pion decays is suppressed above PeV energies, since these
energetic pions are more likely to interact than to decay. 
As a result, while electromagnetic showers are dramatically
stretched in the longitudinal direction, hadronic showers grow much more
slowly with energy. The immediate consequence is that, for a fixed observation
angle away from the Cherenkov cone, the cut-off frequency 
is significantly smaller in electromagnetic than in hadronic
showers. This is apparent in Fig.~\ref{fig:comp-e-p}: At 100 PeV and 
$10^\circ$ inside the Cherenkov cone, the cut-off 
frequency is $\sim 50$ MHz in the electron-induced showers, while it is
$\sim 80$ MHz in the proton-induced ones.  
When EeV energies are reached, this effect becomes very strong. Moreover,
shower-to-shower fluctuations of the longitudinal spread become very large 
due to the LPM effect 
in electron-induced showers \cite{Stanev_LPM}. As a result, some individual
electromagnetic showers have 
a cut-off frequency even 10 times smaller than those in hadronic showers.  
In Fig.~\ref{fig:comp-e-p}, the vertical bars represent the RMS 
of the average of 20 showers, making apparent that
the RMS is much larger in electromagnetic than in hadronic showers
for angles away from the Cherenkov one, for essentially all frequencies. 
It can also be appreciated that fluctuations significantly increase as 
shower energy increases in the case of electron-induced showers, while 
in hadronic showers the dependence on energy is weak. 

\begin{figure}
\begin{center}
\scalebox{0.65}{
\includegraphics{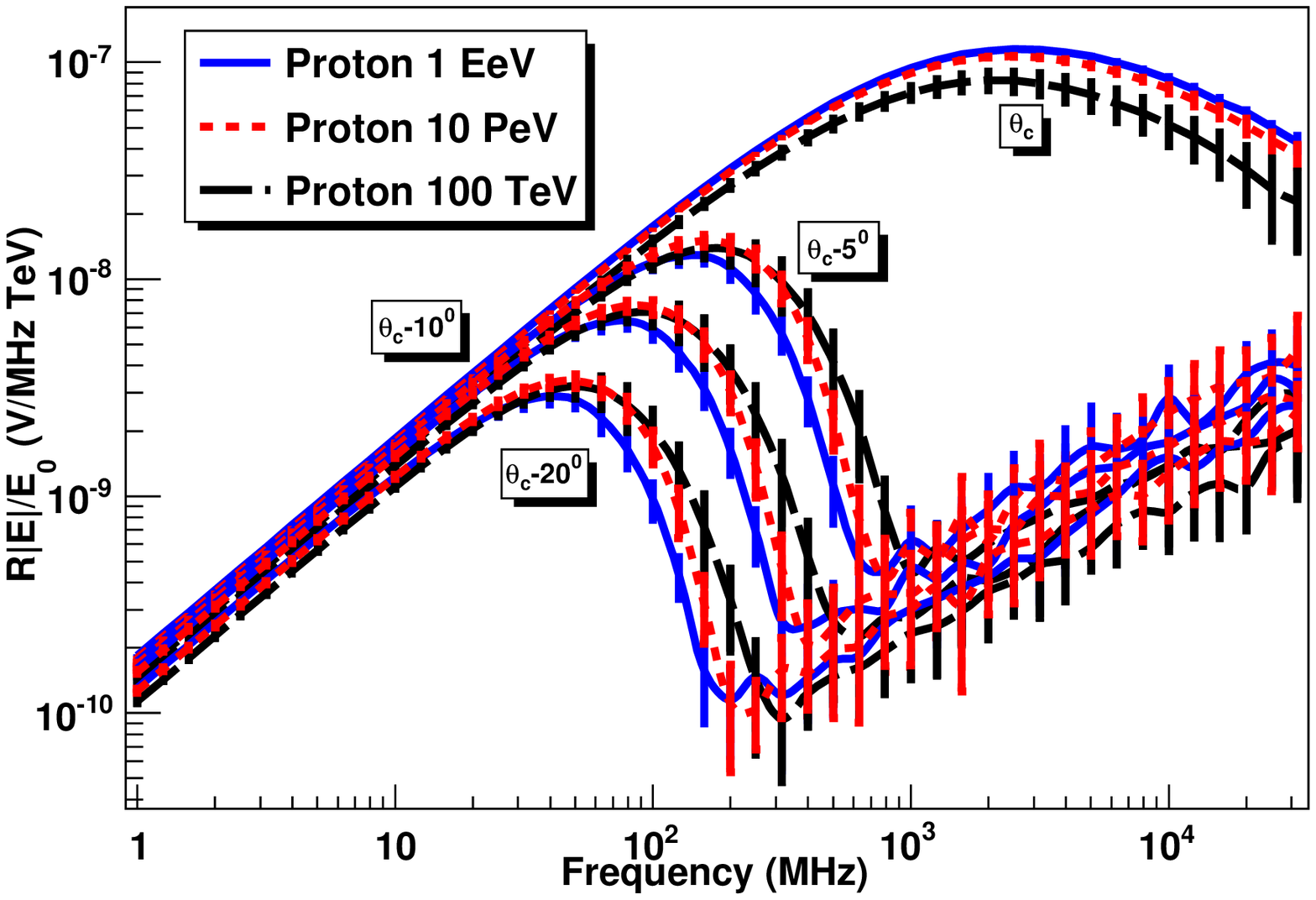}}
\caption{Same as Fig.~\ref{fig:comp-e-p}, but showing only
proton-induced showers at energies $E_0=$100 TeV(long-dashed black line), $10$
PeV (dashed red line) and $1$ EeV (solid blue line). The electric field was divided by shower energy. }
\label{fig:REdivE0}
\end{center}
\end{figure}

We compare hadronic showers among themselves in Fig.~\ref{fig:REdivE0}, where we have divided the electric field by
shower energy to make the differences between the 
features of the spectrum more apparent at different energies. The cut-off frequency at the Cherenkov angle increases slowly as the shower energy rises, since it is determined by the overall lateral spread of the shower, which becomes slightly narrower as the energy increases.
Away from the Cherenkov angle, one can clearly see that the cut-off frequency
decreases with energy as expected, but not so rapidly as in electromagnetic showers, because in hadronic showers the longitudinal spread increases only logarithmically with energy. 

\begin{figure}
\begin{center}
\scalebox{0.65}{
\includegraphics{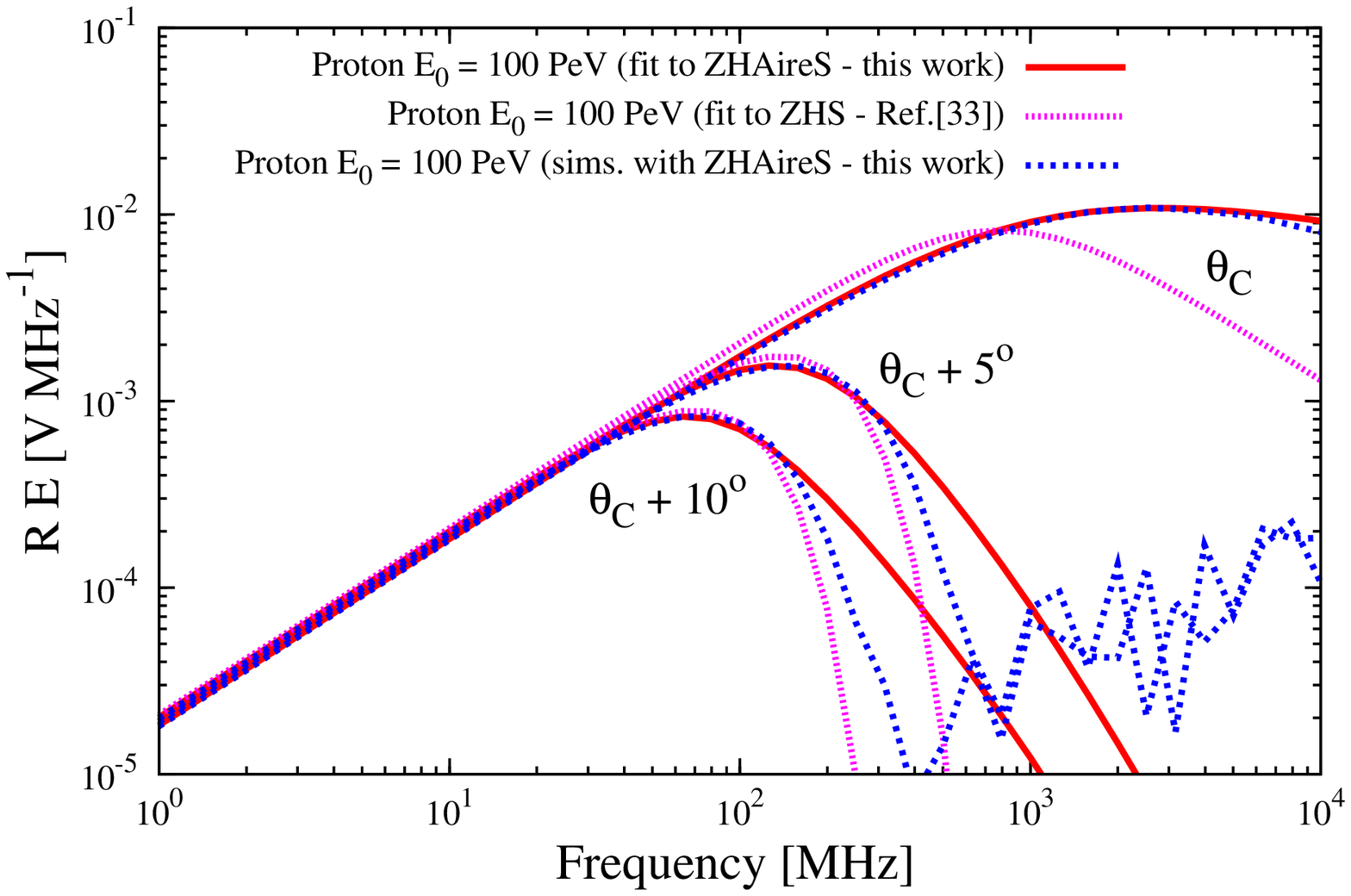}}
\vskip 1cm
\scalebox{0.65}{
\includegraphics{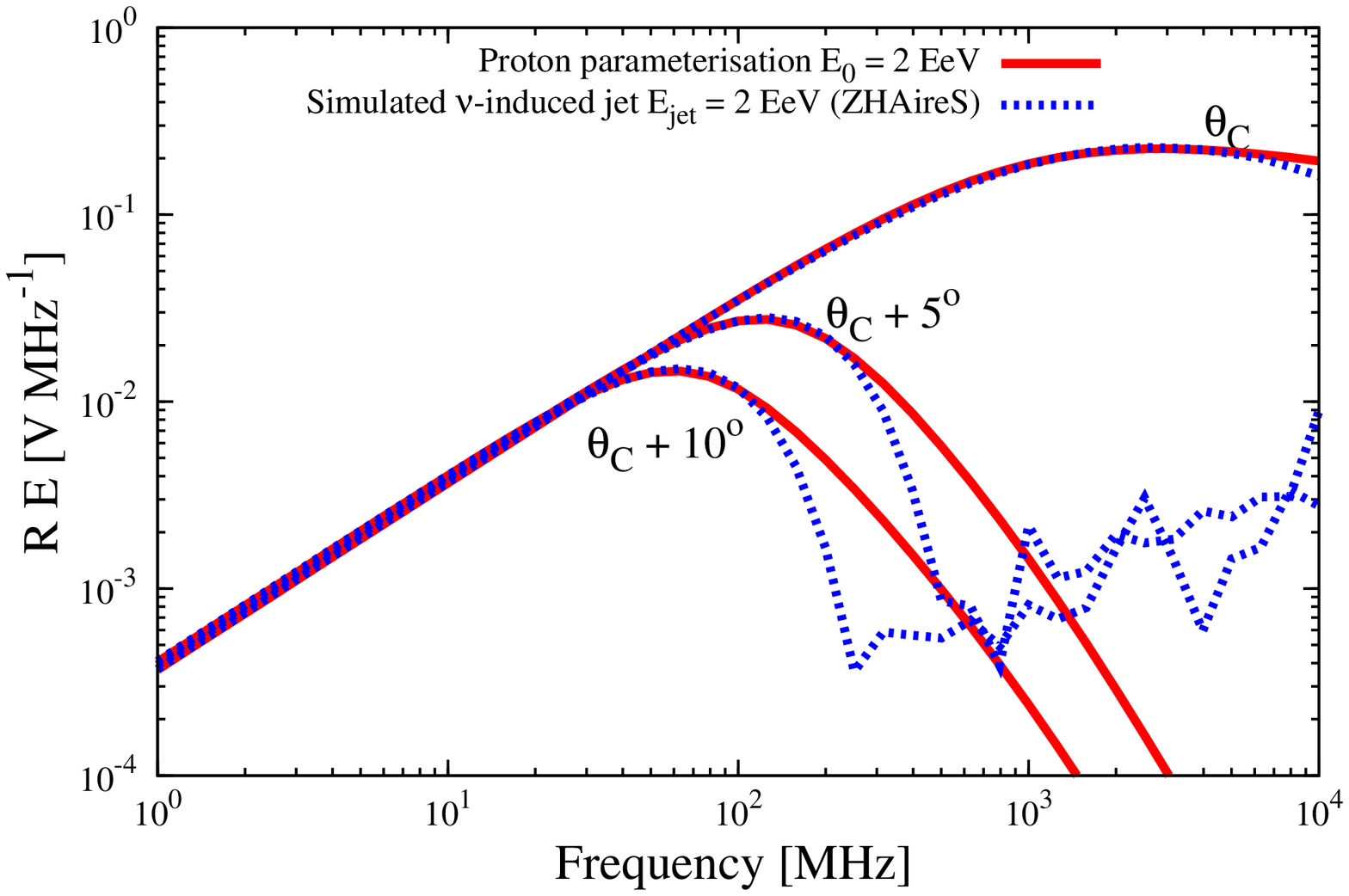}}
\caption{Top panel: 
Frequency spectrum of the Cherenkov radiation obtained
in \ZHAireS simulations of a proton shower with primary 
energy $E_0=100$ PeV (dashed blue line) at the Cherenkov angle and 
$5^\circ$ and $10^\circ$ away from it. The numerical results are compared
to the parameterisations for proton-induced showers  given in Appendix A
(solid  red lines) and in~\cite{alz98} (dotted magenta line). 
Bottom panel: Frequency spectrum obtained in \ZHAireS simulations 
of a neutrino-induced shower in a neutral current interaction (or a charged
current interaction of a muon neutrino) in which the secondaries from the 
fragmentation of the nucleon carry an energy $E_{\rm jet}=2$ EeV (dashed lines). The frequency 
spectrum is compared to the parameteresation of a proton shower with 
energy $E_0=2$ EeV (solid line) at different observation angles. 
See Section \ref{conclusions} for further details.  
}
\label{fig:MC_vs_param}
\end{center}
\end{figure}

In Appendix A we give parameterizations of the frequency 
spectrum of the Cherenkov electric field emitted in hadronic
showers, for practical applications, and we discuss 
their range of validity. The magnitude of the 
electric field is given as a function of proton
energy, observation angle and frequency. 
As an example we show in Fig.~\ref{fig:MC_vs_param} the electric 
field spectrum at the Cherenkov angle in a 100 PeV proton shower,
compared to the corresponding parameterization as given in 
Appendix A. As stated in Appendix A, one can see that the 
parameterizations work very well (accuracy $\sim 1\%$) for frequencies up to 
the frequency $\nu_{\rm max}$, at which the spectrum is maximum for each
observation angle. For frequencies above $\nu_{\rm max}$, the accuracy
worsens gradually and reaches $\sim 5\%$ at $\nu=2\nu_{\rm max}$,
for observation at the Cherenkov angle $\theta=\theta_C$, and $\sim 15\%$ at
angles $\theta=\theta_C\pm 10^\circ$. In the top panel of
Fig~\ref{fig:MC_vs_param} we also show parameterizations of the frequency
spectrum of 100 PeV proton-induced showers as obtained in~\cite{alz98}. Those
parameterizations are based on 1-dimensional simulations and the lateral
distribution was only acccounted for in a very approximate way, which explains
the large discrepancies with the results of this work at the Cherenkov
angle. However, the fits are in fairly good agreement for angles away from the
Cherenkov angle since the features of the spectrum at those angles are
determined by the longitudinal development of the shower which is correctly accounted for in~\cite{alz98} as well as in this work.

\subsection{Contribution of charged pions, muons and protons to excess track lengths.}

In a hadronic shower, there are more particles besides electrons
that can be expected to contribute to the excess charge. In fact, it has been 
experimentally determined \cite{MINOS} in atmospheric showers that cascades 
develop an excess of positive muons over negative muons, stemming from an excess of positively charged pions and kaons \cite{gaisser}
. As a result, we expect pions and muons to contribute with an excess of positive track lengths to the Cherenkov radio emission. 

\begin{figure}
\begin{center}
\scalebox{0.65}{
\includegraphics{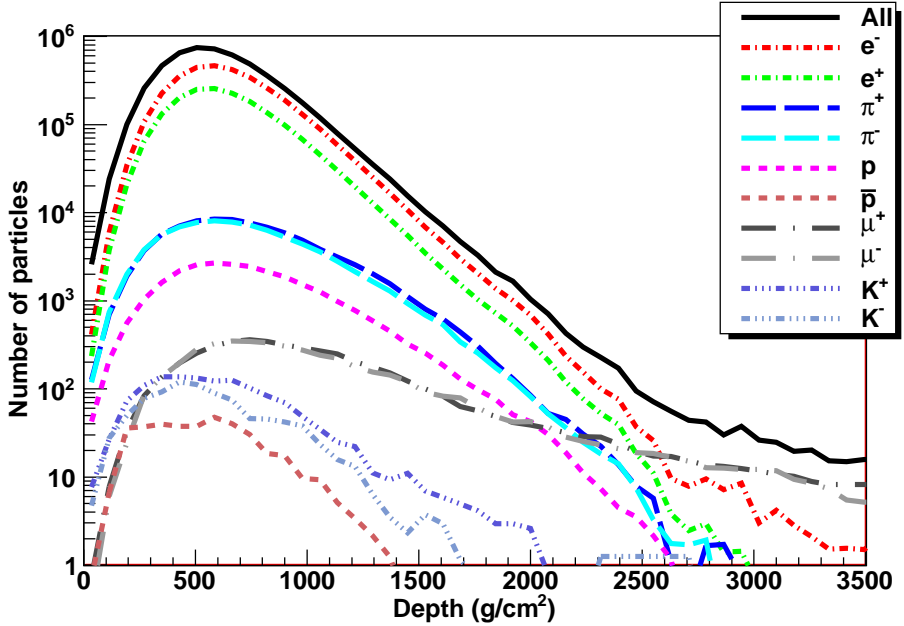}}
\caption{Longitudinal development of of electrons, positrons, protons,
  anti-protons, $\pi^+$, $\pi^-$, $\mu^+$, $\mu^-$, $K^+$ and $K^-$ as
  obtained in \ZHAireS simulations of proton-induced showers in ice at 1 PeV.}
\label{fig:long}
\end{center}
\end{figure}

In Fig.~\ref{fig:long} we show the average longitudinal profile of electrons, positrons, protons, anti-protons, 
$\pi^+$, $\pi^-$, $\mu^+$, $\mu^-$, $K^+$ and $K^-$ as obtained in \ZHAireS
simulations of 20 proton-induced showers of $E_0=1$ PeV in ice. It becomes
apparent that the main contributions to the total projected tracklength, which
is approximately proportional to the area under the longitudinal profile, come
from charged pions, protons and muons. We neglect the contribution from
charged kaons, which amounts to $\sim 1\%$ of that due to pions. Protons
contribute most to the excess positive tracklength because almost no anti-protons are produced in the shower to compensate for them, as can be seen in Fig.~\ref{fig:long}.

In order to quantify the influence of charged pions, muons and protons on the
track lengths, we computed both, the total and projected track lengths of those particles in \ZHAireS simulations.
As expected, we found that there is an excess of track lengths due to 
positively charged pions, muons and protons. For example, 
in a 10 PeV proton shower:

\begin{equation}
R_T = \frac{(T_{\pi^+}-T_{\pi^-})+(T_{\mu^+}-T_{\mu^-})+(T_p-T_{\bar p})}{(T_{e^-}-T_{e^+})} \sim 0.015\;,  
\label{ratio}
\end{equation} 

\noindent where $T$ denotes the projected track length due to different particles. 
This means that the excess of positively charged 
pions, muons and protons diminishes the excess projected track length of the shower 
by $\sim 1.5\%$ because the excess is of opposite sign, compared to the excess 
of electrons. This contribution decreases in magnitude with shower energy in 
the same fashion as the EM energy ratio increases (see Fig.~\ref{fig:mckay}
and Table \ref{tab:tracks_p}). 
Also, for a fixed shower energy, there is a dependence on the hadronic 
interaction model used in the simulations, as will be discussed in the 
next section.

In the top half of table~\ref{tab:tracks_p} we show the total and projected track length divided by the primary
energy $E_0$ due to different particles in proton-induced showers of $E_0=$1
TeV, 1 PeV and 1 EeV. Note
that the $e^{\pm}$ track lengths normalized to $E_0$ increase with energy, since the
energy feeding the electromagnetic component of the shower also increases with
the energy of the primary proton. This is illustrated by the EM energy
fractions shown in the bottom of table~\ref{tab:tracks_p}. As discussed
before, this increase in the EM energy ratio is what causes the proton-induced
shower projected excess $e^{\pm}$ tracks to increase with energy, as
can be seen in figure \ref{fig:mckay}. Furthermore, both the total $e^{\pm}$
track/$E_0$ and the projected $e^{\pm}$ track/$E_0$  are proportional to the EM energy fraction of the
shower. The ratio between the total $e^{\pm}$ track length normalized to
$E_0$, and the EM energy fraction of the shower is constant,
i.e. $(\sum{T_{e^{\pm}}}/E_0)/EM^{fraction}\simeq 57$ m/TeV. The
equivalent ratio for the projected $e^{\pm}$ tracks varies only slightly
between 41 and 45 m/TeV for all the energy range shown in table \ref{tab:tracks_p}.

\begin{table*}[!htb]
\begin{center}
\begin{tabular}{ c | c c c || c c c } \hline
\multicolumn{4}{c||}{Total Track/$E_{0}$ [m/TeV] } &\multicolumn{3}{c}{Proj. Track/$E_{0}$ [m/TeV]}\\ \hline\hline
        & 1 TeV & 1 PeV & 1 EeV & 1 TeV & 1 PeV & 1 EeV\\\hline
  $e^+$     & 1271(145) & 1679(43)  & 1747(16)   & 1024(158) & 1466(47)  & 1539(17)\\
  $e^-$     & 2629(287) & 3440(86)  & 3573(32)   & 1801(262) & 2544(79)  & 2664(30)\\
  $\pi^+$   & 280(69)   & 93(20)    & 62(7)      & 195(50)   & 65(14)    & 43.1(5.2)\\  
  $\pi^-$   & 277(74)   & 90(20)    & 60(7)      & 184(51)   & 59(13)    & 39.6(4.8)\\ 
  $\mu^+$   & 14.7(7.3) & 5.1(1.2)  & 3.6(0.6)   & 11.0(6.8) & 3.8(0.9)  & 2.4(0.4)\\  
  $\mu^-$   & 14.4(7.2) & 5.0(1.5)  & 3.3(0.4)   & 9.6(6.1)  & 3.6(1.3)  & 2.16(0.35)\\  
  $p$       & 85(24)    & 27.9(6.3) & 18.6(2.2)  & 68(20)    & 21.6(4.9) & 14.4(1.7)\\  
  $\bar p$  & 0.4(0.9)  & 0.3(0.2)  & 0.19(0.03) & 0.3(0.9)  & 0.3(0.2)  & 0.18(0.03)\\  \hline\hline

\end{tabular}

\vspace{0.6cm}

\begin{tabular}{ c | c || c} \hline

& $EM^{fraction}$ proton-induced & $EM^{fraction}$ electron-induced \\ \hline\hline
1 TeV & 68.7\% & 99.6\% \\ \hline
1 PeV & 89.2\% & 99.2\% \\ \hline
1 EeV & 92.6\% & 98.0\% \\ \hline\hline
\end{tabular}
\end{center}
\caption{Top table: Average total track lengths (top left) and projected track lengths
  (top right) divided by $E_0$ for different particle species obtained from
  simulations of  proton-induced showers of $E_0=$1 TeV, 1 PeV and 1 EeV in
  ice, using \zhaires. Also shown in parenthesis are the RMS values for the 20
  showers. Bottom table: Fraction of shower energy feeding the EM component for proton- and electron- induced
  showers of $E_0=$1 TeV, 1 PeV and 1 EeV. The cuts used in these simulations
  were 80 keV for $e^\pm$ and 500 keV for hadrons.}
\label{tab:tracks_p}
\end{table*}

\begin{figure}
\begin{center}
\scalebox{0.65}{
\includegraphics{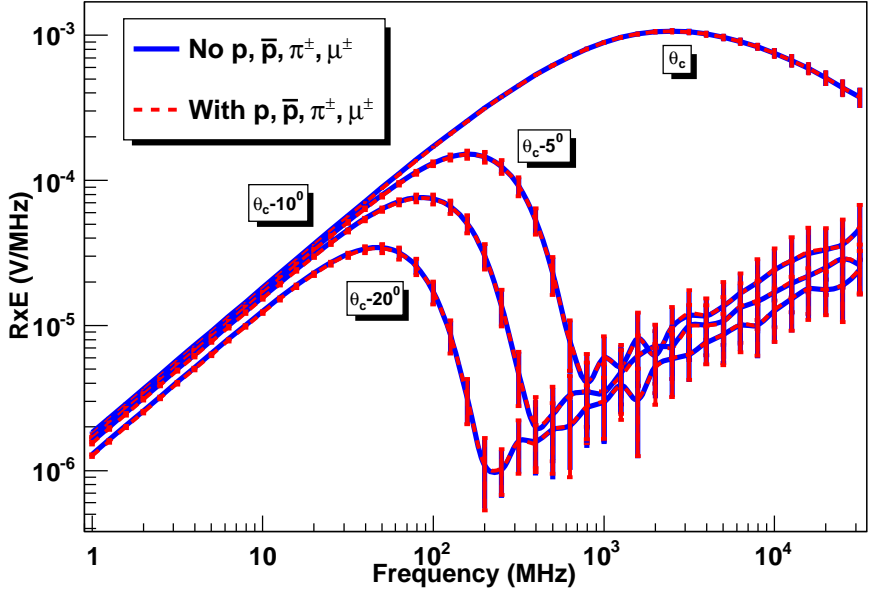}}
\caption{
Average frequency spectrum of the Cherenkov radiation obtained
from \ZHAireS simulations of $10$ PeV proton-induced showers accounting (red dashed lines) and not 
accounting (blue solid line) for pions, muons and protons in the 
calculation of the electric field.
The spectrum is shown at four observation 
angles with respect to shower axis. The RMS of the 20 simulated showers 
is also shown.
}
\label{fig:comp-pi-nopi}
\end{center}
\end{figure}

The very small contribution of pions,  muons and protons to the field can be seen in Fig.~\ref{fig:comp-pi-nopi}, where we show the
Fourier-spectrum of the electric field in proton showers of 10 PeV with and without
accounting for the charged pions, muons and protons tracks. Although the
actual field emitted by the hadrons in the shower is very small, compared to
the field emitted by electrons and positrons, the hadronic component carries a
significant fraction of the shower energy, especially at lower energies. The
importance of this energy balance of the shower can be illustrated by the
effects of the EM energy fraction on the radio emission, as discussed
before. Given our results, following the actual tracks of hadrons in the shower turns out not to be essential for the field calculation, but accounting for the
hadronic energy of the shower is.  

%%%%%%%%%%%%%%%%%%%%%%%%%%%%%%%%%%%%%%%%%%
\subsection{Hadronic interaction model dependence}

\begin{figure}
\begin{center}
\scalebox{0.65}{
\includegraphics{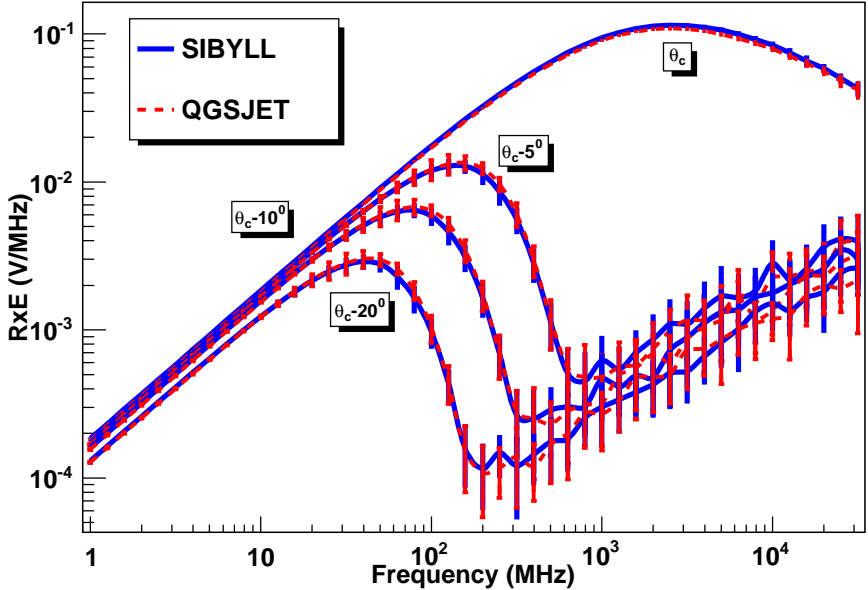}}
\caption{
Average frequency spectrum of the Cherenkov radiation obtained
from \ZHAireS simulations of $1$ EeV proton-induced showers using QGSJET01 
(red dashed lines) and SIBYLL 2.1 (blue solid line).  
The spectrum is shown at four observation 
angles with respect to shower axis. The RMS of the 20 simulated showers 
is also shown.
}
\label{fig:models}
\end{center}
\end{figure}

Modeling of hadronic showers above PeV energies 
relies on the extrapolation of hadronic interactions
to energies and regions in the parameter space of the 
collisions which have not been probed in terrestrial 
accelerators \cite{hadronic_models}. We have studied 
the influence of the hadronic model on the coherent Cherenkov radiation
frequency spectrum. For that purpose we have simulated 
a set of proton-induced showers with the QGSJET01 model \cite{QGSJET01},
and compared the spectrum with that obtained using the
SIBYLL 2.1 model \cite{SIBYLL21}, which we have been using by default in all
other simulations so far\footnote{We use the SIBYLL 2.1 and QGSJET01 models 
since recent data favors them over other models \cite{horandel09}.}. 
This comparison is shown in Fig.~\ref{fig:models}. The differences
between the normalization of the spectra in the coherent region obtained with
SIBYLL 2.1 and QGSJET01 are small, with the QGSJET01 spectrum smaller
than that predicted with SIBYLL 2.1 by up to $\sim 4\%$ around the
  cutoff frequency at the Cherenkov angle.

%It is well-known that QGSJET01 predicts more muons and pions per interaction
%than SIBYLL 2.1 \cite{ostapchenko}. For instance, we have obtained the
%longitudinal profile of the number of pions in proton-induced showers, and
%observed that the number of pions at the maximum is $\sim 24\%$ larger in
%QGSJET01 than in SIBYLL 2.1. As a consequence of this, we expect a difference
%between models in the relative contribution of the excess of positive hadrons
%to the excess track length in the shower, quantified by the ratio $R_T$ - see
%Eq.~(\ref{ratio}). While in SIBYLL 2.1 $R_T\sim 0.011$ at 1 EeV, QGSJET01
%predicts a larger ratio, $R_T\sim 0.013$. Also, more pions, protons and muons
%means less electrons and positrons in the shower, decreasing the negative
%excess track, which is made apparent by the decrease of the EM energy ratio
%from 93\% using SYBILL to 91\% when using QGSJET. This is consistent with the
%fact that the electric field in the fully coherent region is slightly smaller
%when using QGSJET01 than when SIBYLL 2.1 is adopted.
 
It is well known that QGSJET01 produces more muons and pions per interaction
than SIBYLL 2.1~\cite{ostapchenko} and also protons. For instance, we have
obtained the longitudinal profile of the number of pions in proton-induced
showers, and observed that the number of pions at the maximum is $\sim 24\%$
larger in QGSJET01 than in SIBYLL 2.1~\cite{ostapchenko}. Also more protons,
pions, and muons per interaction means less electrons and positrons in the
shower, decreasing the negative excess track (denominator of Eq. (\ref{ratio})
). As a consequence, we expect a difference in the ratio $R_T$ predicted by
both models. While in SIBYLL 2.1 $R_T \sim 0.012$ at 1 EeV, QGSJET01 predicts
a larger ratio $R_T \sim 0.014$, the increase being dominated by the proton projected track-length. Also as a consequence, the EM energy ratio is larger (93\%) in SIBYLL when compared to QGSJET (91\%) at 1 EeV. This is consistent with the fact that the electric field in the fully coherent region is slightly smaller when using QGSJET01 than when SIBYLL 2.1 is adopted. We remark here that the smallness of $R_T$ reflects that the tracks of protons, pions and muons are mostly irrelevant to the field calculation, and this conclusion holds regardless of the details of the high-energy hadronic interaction model. In fact, we have also studied the effect of the low energy hadronic models on the radio emission, and our conclusions remain the same.

There are also differences in the frequency at which the electric   field intensity is maximum at each observation angle away from the Cherenkov angle. These stem from the fact that the longitudinal profile of the excess negative charge predicted by SIBYLL 2.1 is longer than in QGSJET01, leading to higher frequencies at the maximum when using QGSJET01. Also, the lateral profile is slightly flatter in QGSJET, leading to a slightly smaller cut-off frequency at the Cherenkov angle. However, these differences are small. 

%Model dependence and photo-production in EM showers ??. 

%%%%%%%%%%%%%%%%%%%%%%%%%%%%%%%%%%%%%%%%%%
\section{Summary and conclusions}
\label{conclusions}

We have presented \zhaires, a Monte Carlo code that merges the high energy
hadronic interaction and tracking capabilities of AIRES\cite{aires}, 
and the dense media propagation capabilities of TIERRAS \cite{TIERRAS} with 
the precise low energy $e^\pm$ tracking and radiation calculation 
capabilities of ZHS\cite{ZHS91,ZHS92}. This combination allows a  
precise full simulation of radio emission of electromagnetic,
and for the first time also hadronic showers up to EeV energies. \ZHAireS has been used in 
conjuction with the TIERRAS package \cite{TIERRAS} to simulate showers in 
dense media, obtaining the radiation emitted due to the Askar'yan effect.

We have compared \ZHAireS results for electromagnetic showers in ice against
ZHS up to 10 EeV, obtaining a good agreement between them. In the case of hadronic showers, we have compared the 
results from \ZHAireS and GEANT4 up to 90 TeV \cite{McKay_radio}, the highest 
energy for full simulations of hadronic showers with GEANT in the literature, also with very good
quantitative agreement. 
Unlike GEANT4, \ZHAireS is capable of simulating the emission from
hadronic showers up to EeV energies.

By comparing the results of thinned \ZHAireS simulations with un-thinned ones
up to 1 PeV, we obtained thinning parameters which give a good compromise
between accuracy and CPU time. 
We then tested the
applicability of these parameters at higher energies, by comparing our results
against ZHS, which has its own, well established \cite{aljpz09}, thinning 
algorithms and parameters.

We confirmed the expectation that the EM energy ratio of hadronic showers keeps
approaching that of EM showers at the highest energies, increasing the
magnitude of the electric field (normalized by $E_0$) as energy grows. We also showed that, in the
case of UHE purely EM showers, the EM energy ratio deviates from a linear dependence on
energy, due to photohadronic interactions. 

By comparing EM and hadronic showers up to 10 EeV we have found that, at the
Cherenkov angle, the cut-off frequency tends to be smaller in hadronic
showers and increases slowly as the energy rises because the lateral
distribution becomes narrower. For angles away from the
Cherenkov angle, as expected, the cut-off frequency decreases with energy, due
to the logarithmic growth of the longitudinal profile.
In contrast, away from the Cherenkov angle, the cut-off frequency for EM 
showers
decreases rapidly above PeV energies due to the LPM effect, which is much
more pronounced in EM showers than in hadronic ones. We also found that 
the fluctuations in hadronic
showers are almost independent of energy, while in EM showers they grow
rapidly, as the LPM effect becomes important.

We have also analyzed the influence of charged pions, muons and protons on the radio
emission of showers. We found that most of their contribution is due
to the excess of protons, with pions being the next contribution
in importance, that induce a decrease of the (total) negative
excess track length of the shower of $\sim1-2\%$ above 1 PeV. Although the
actual field emitted by the hadrons in the shower is very small, the hadronic component carries a
significant fraction of the shower energy. The
importance of this energy balance is illustrated by the
effects of the EM energy fraction on the radio emission. Given our results, following the actual tracks of hadrons in the shower turns out not to be essential for the field calculation, but accounting for the
hadronic energy of the shower is.

We have compared the results obtained using SIBYLL 2.1 \cite{SIBYLL21} and 
QGSJET01 \cite{QGSJET01}, and found that
there is a small dependence of the normalization of the emitted 
electric field on the hadronic model, up to $\sim 4\%$ lower in QGSJET01 at
the Cherenkov angle. This is due to the higher number of protons, pions and
muons when QGSJET01 is used, which increases the positive projected excess
track-length of protons, pions and muons, with protons dominating the
difference. Also more protons, pions and muons means means less electrons and
positrons in the shower, decreasing the negative excess track. The net effect
is a lower negative projected excess track as a whole, diminishing the
field. At angles away from the Cherenkov angle, the longitudinal excess
profile becomes important, and the shorter showers obtained using QGSJET lead
to the maximum of the emission at higher frequencies, when compared to SIBYLL showers.

Finally, we have used our simulations of the emission due to the Askar'yan 
effect in hadronic showers to obtain new parameterizations of the radio pulse 
frequency spectrum which are described in Appendix A. 
We note that our results for proton 
showers can also be applied to approximately model the Cherenkov emission 
from hadronic showers induced in high-energy neutrino interactions.
This is shown in the bottom panel of Fig.~\ref{fig:MC_vs_param}
where we have plotted the frequency spectrum obtained in \ZHAireS simulations 
of a neutrino-induced shower in a neutral current interaction (or a charged
current interaction of a muon neutrino) in which the secondaries from the 
fragmentation of the nucleon carry 2 EeV of energy in total. The products
of the neutrino-nucleon
interaction, mainly pions, were obtained with the HERWIG interaction Monte Carlo code \cite{HERWIG}. 
The frequency 
spectrum obtained in \ZHAireS simulations is compared to the parameterization of a proton shower 
with energy $E_0=2$ EeV as given in Appendix A.

%%%%%%%%%%%%%%%%%%%%%%%%%%%%%%%%%%%%%%%%%%
\section{Acknowledgments}

J.A-M, W.R.C. and E.Z. thank Xunta de Galicia (INCITE09 206 336 PR) and 
Conseller\'\i a de Educaci\'on (Grupos de Referencia Competitivos -- 
Consolider Xunta de Galicia 2006/51); Ministerio de Ciencia e Innovaci\'on 
(FPA 2007-65114, FPA 2008-01177 and Consolider CPAN) and Feder Funds, Spain.
We thank CESGA (Centro de SuperComputaci\'on de Galicia) for computing resources
and assistance. 
M.T. thanks Fundaci\'on Universidad Nacional de Cuyo 
and CONICET (Argentina) for financial support. 
 We thank J. Bray, R. Crocker, C.W. James, 
R.J. Protheroe, A. Romero-Wolf and R.A. V\'azquez for many helpful discussions. 
Special thanks to Gonzalo Rodr\'\i guez-Fern\'andez 
for help with GEANT4 simulations.

\section{Appendix A}
\label{appendix}
 
In this Appendix we give parameterizations of the magnitude
of the Cherenkov electric field in proton-induced showers in ice
for practical appications,
as obtained in simulations with \zhaires.

This is the first time parameterizations of the electric field in hadronic
 showers are presented. The parameterizations given here complement
those presented in \cite{aljpz09} for purely electromagnetic showers. 
  However in \cite{aljpz09} photoproduction interactions were not 
taken into account, and for this reason the parameterizations presented
there for electromagnetic showers should be less accurate for $E>10^{19}$ eV.
This important subject will be further studied in a future work.

The physical basis for our parameterization of the radiated spectrum is 
the ``box model" of shower development \cite{ZHS92,buniy02,alvz06,aljpz09}. 
In this model, the distribution of particle tracks  
has a characteristic length $L$, proportional to the 
radiation length $X_0$, and a width $R$, proportional to the 
Moli\`ere radius $R_M$. 
The model motivated a functional form for the radiated spectrum $\vec{E}(\nu)$
in electromagnetic showers, given by \cite{aljpz09}: 
\begin{eqnarray}
r |\vec{E}(E_0,\theta,\nu)| & = 
& A(E_0,\theta,\nu) \, \times \, d_L(E_0,\theta,\nu) \, \times \, d_R(E_0,\theta,\nu) 
\label{functionalform}
\end{eqnarray}
where $A$ is a coherent amplitude that increases linearly with 
frequency, which is multiplied by lateral and longitudinal 
decoherence factors $d_R$ and $d_L$, defined in such a way so that 
$d_{R,L} \le 1$ and $d_{R,L}\rightarrow 1$ as $\nu \rightarrow 0$. 
Also, as $\theta\rightarrow\theta_C$, where $\theta_C$ is the Cherenkov angle, $d_L\rightarrow 1$, since interference
over the width of the shower dominates.
In general, $A$, $d_L$ and $d_R$ depend on the shower energy 
$E_0$, frequency $\nu$, and observation angle $\theta$ with respect to the shower axis. 
$r$ is the (far-field) observation distance. 

In this paper we adopt this same functional form for the radiated spectra
in proton-induced showers. For the decoherence factors we use the same
functional forms as in \cite{alvz06}, namely:
\begin{eqnarray}
d_{R [L]} & = & \frac{1}{1+(\nu/\nu_{R [L]})^{\bar{\alpha} [\bar{\beta]}}}. \label{drl}
\end{eqnarray}
where $\nu_R$ and $\nu_L$ are characteristic frequencies at which lateral and longitudinal 
decoherence become important, and $\bar{\alpha}$ and $\bar{\beta}$ give the strength of the decoherence. 
The frequencies $\nu_L$ and $\nu_R$ are inversely proportional to $L$ and $R$,
respectively. 
The amplitude $A$ of the fully coherent (low-frequency) component is proportional to the total excess track length $T$, which in hadronic showers is known to
deviate from a linear behavior with shower energy (see Fig.~\ref{fig:mckay}). 
The length $L$, width $R$, and excess track length $T$ of the shower can 
be related to properties of the interaction medium \cite{ZHS92,buniy02,alz97,alvz06}:  
\begin{eqnarray}
\nu_L(\theta) ~&~ \approx ~&~ \frac{c}{L} \frac{1}{| 1-n \cos \theta |} ~=~ \frac{\rho}{\bar{k}_L X_0} \frac{c}{|1-n \cos \theta|} \;,\label{kleqn}\\
\nu_R (\theta = \theta_C) ~&~ \approx ~&~ \frac{c/n}{R} ~=~ \frac{\rho}{\bar{k}_R R_M} \frac{c}{\sqrt{n^2-1}} \;, \label{kreqn}\\
A(E_0,\theta,\nu) ~&~ = ~&~ \bar{k}_E~\nu~T \sin \theta ~\approx~ \bar{k}_E \frac{E_0}{E_C} \frac{X_0}{\rho} \nu_{\rm MHz} \sin \theta  \;,\label{keeqn}
\end{eqnarray}
where $\nu_{\rm MHz}$ is the frequency in MHz,
${\bar k}_E$ has units of ${\rm V~cm^{-1}~MHz^{-2}}$ and
 $c=3\cdot10^{10}~{\rm cm~s^{-1}}$ is the speed of light.
$X_0=36.08~{\rm g~cm^{-2}}$, $R_M=10.57~{\rm g~cm^{-2}}$, $n=1.78$, 
$\rho=0.924~{\rm g~cm^{-3}}$ and $E_C=73.1~{\rm MeV}$ are respectively 
the radiation length, Moli\`ere radius, refractive index, density, 
and critical energy of ice.  
Throughout, we use the approximation $\nu_R (\theta) = \nu_R(\theta_C)$,
since lateral decoherence is only important near the Cherenkov angle. 

The quantity $\bar{k}_E$ is also energy dependent, since
the track length due to the excess negative charge deviates
from linearity in hadronic showers (see Fig.~\ref{fig:mckay}). 
To fit $\bar{k}_E (E_0)$ we use:

\begin{equation}
\bar{k}_E = k_{E,0}~\tanh \left(\frac{\log_{10}E_0 - \log_{10}E_E}{k_{E,1}}\right) \;,
\label{fit_kE}
\end{equation}
where $k_{E,0}$, $k_{E,1}$ and $E_E$ are constants given in Table~\ref{param_table_kE}.

Since $L$ is dependent upon the shower energy $E_0$, $\bar{k}_L$ is in fact 
also energy-dependent, although with a much weaker dependence than in electromagnetic
showers, because hadronic showers are not so strongly affected by the LPM effect. 
For this reason we expect $\bar{k}_L$ to be well fit in the whole energy range 
by the relation:
\begin{equation}
\bar{k}_L  =  k_{L,0}~\left(\frac{E}{E_L}\right)^{\gamma} \;,
\label{fit_kL}
\end{equation}
where $k_{L,0}$, $E_L$ and $\gamma$ are constants given in Table~\ref{param_table_kL}. 

We also found a weak dependence of $\bar{k}_R$ with energy, 
which we further parameterize as:
\begin{equation}
\bar{k}_R = k_{R,0} + \tanh \left(\frac{\log_{10}E_R - \log_{10}E_0}{k_{R,1}}\right) \;,
\label{fit_kR}
\end{equation}
where $k_{R,0}$, $k_{R,1}$ and $E_R$ are constants given in Table~\ref{param_table_kR}.

Finally, $\bar{\alpha}$ and $\bar{\beta}$ are found to be
practically independent of shower energy, and they are given 
in Table~\ref{param_table}. 

Our fitting procedure is the same as in \cite{aljpz09}. We reproduce it here for
self-consistency of the paper. For each shower of energy $E_0$, we start by
obtaining $A(\theta,\nu)$ and $d_R$ from a fit to the spectrum at the Cherenkov angle $\theta_C$
where $d_L=1$. 
We then obtain $\bar{k}_E$ by substituting
$A(\theta,\nu)$ in Eq. (\ref{keeqn}). 
Also from the fitted $d_R$ in Eq.(\ref{drl}), we obtain parameters $\nu_R$
and $\bar{\alpha}$, and relate $\nu_R$ to
$\bar{k}_R$ via Eq.\ (\ref{kreqn}). Fixing $A$ and $d_R$ to their fitted values, we then allow
$d_L$ to vary on fits of Eq. (\ref{functionalform}) to the simulated field at
various angles away from the Cherenkov angle, obtaining
$d_L(\theta,\nu)$ for each angle. From these we can obtain $\nu_L(\theta)$ and
$\bar{\beta}$ from Eq. (\ref{drl}), and $\bar{k}_L$ from Eq. (\ref{kleqn}), respectively. We repeat this process many times, varying the energy $E_0$ in multiples of 10 over the energy range from $1$ TeV to $10$ EeV. By using the values
of $\bar{k}_E(E_0)$, $\bar{k}_L(E_0)$ and  $\bar{k}_R(E_0)$ from fits at these different energies, we obtain the
parameters of Eqs.~(\ref{fit_kE}), (\ref{fit_kL}) and (\ref{fit_kR}), that
define the energy dependence of $\bar{k}_E$, $\bar{k}_L$ and  $\bar{k}_R$, and
are given in Tables \ref{param_table_kE}, \ref{param_table_kL} and
\ref{param_table_kR}, respectively. The fitted parameters $\bar{\alpha}$ and
$\bar{\beta}$, practically energy independent, are given in Table \ref{param_table}. 

%%%%%
\begin{table}
\begin{center}
\renewcommand{\arraystretch}{1.9}
\begin{tabular}{ c  c  c }
\hline
$~~~k_{E,0}~{\rm [V~cm^{-1}~MHz^{-2}]}~~~$ & $~~~k_{E,1}~~~$ & $~~~\log_{10}(E_E/{\rm eV})~~~$ \\ \hline 
\hline
$4.13~10^{-16}$ & $2.54$ & $10.60$ \\ \hline
\end{tabular}
\caption{Fitted proton-induced shower parameters in ice, as defined by Eq.\ (\ref{fit_kE}).
}
\label{param_table_kE}
\end{center} 
\end{table}

%%%%%
\begin{table}
\begin{center}
\renewcommand{\arraystretch}{1.9}
\begin{tabular}{  c   c   c  }
\hline
$~~~k_{L,0}~~~$ & $~~~\gamma~~~$ & $~~~\log_{10}(E_L/{\rm eV})~~~$ \\ \hline
\hline
$31.25$ & $3.01~10^{-2}$ & $15.00$ \\ \hline
\end{tabular}
\caption{Fitted proton-induced shower parameters in ice, as defined by Eq.\ (\ref{fit_kL}).
}
\label{param_table_kL}
\end{center} 
\end{table}

%%%%%
\begin{table}
\begin{center}
\renewcommand{\arraystretch}{1.9}
\begin{tabular}{  c   c   c  }
\hline
$~~~k_{R,0}~~~$ & $~~~k_{R,1}~~~$ & $~~~\log_{10}(E_R/{\rm eV})~~~$ \\ \hline
\hline
$2.73$ & $1.72$ & $12.92$ \\ \hline
\end{tabular}
\caption{Fitted proton-induced shower parameters in ice, as defined by Eq.\ (\ref{fit_kR}).
}
\label{param_table_kR}
\end{center} 
\end{table}

%%%%%
\begin{table}
\begin{center}
\renewcommand{\arraystretch}{1.9}
\begin{tabular}{  c   c  }
\hline
$~~~~\bar{\alpha}~~~~$ & $~~~~\bar{\beta}~~~~$ \\ \hline
\hline
$1.27$ & $2.57$  \\ \hline
\end{tabular}
\caption{Fitted proton-induced shower parameters in ice, as defined by Eq.\ (\ref{drl}).
}
\label{param_table}
\end{center} 
\end{table}

While all the fitted parameters vary on a shower-to-shower basis, variations tended to be small, so that
approximating $k_E \approx \bar{k}_E$, $\beta \approx \bar{\beta}$, etc., is appropriate for individual showers.
This is in contrast to previous parameterizations of electromagnetic showers, in which $k_L$ 
varied strongly from the mean fitted values $\bar{k}_L$, due to variations in the longitudinal
spread of the showers caused by the LPM effect. This is not the case in hadronic showers and
we take $k_L \approx \bar{k}_L$.

Finally, the fits have an accuracy of $\sim 1\%$ for frequencies up to 
the frequency $\nu_{\rm max}$, at which the spectrum is maximum for each 
observation angle. For frequencies above $\nu_{\rm max}$, the accuracy
worsens gradually and reaches $\sim 5\%$ at $\nu=2\nu_{\rm max}$,
for observation at the Cherenkov angle $\theta=\theta_C$, and $\sim 15\%$ at angles 
$\theta=\theta_C\pm 10^\circ$.  

\end{document}